\documentclass[a4paper,10pt,openany]{article}
\usepackage{setspace}
\usepackage{hyperref}
\usepackage{pst-3d}
\usepackage{pst-3dplot}
\usepackage[utf8]{inputenc}
\usepackage[english]{babel}
\usepackage{amsmath,amssymb,amsthm,mathrsfs,amsfonts,dsfont}
\usepackage{graphicx}
\usepackage[small, skip=10pt]{caption}
\usepackage{epigraph}
\usepackage{indentfirst}
\usepackage{booktabs}
\usepackage{fancyhdr}
\usepackage{vmargin}
\usepackage{feynmf}
\usepackage{yfonts}
\usepackage{lscape}
\usepackage{textcomp}
\usepackage{pstricks}
\usepackage[metapost]{mfpic}
\usepackage{fancybox}
\usepackage{slashed}
\usepackage[verbose]{wrapfig}
\usepackage{pifont}
\usepackage{keystroke}
\usepackage{appendix}
\usepackage{enumitem}
\usepackage{array}
\usepackage{colortbl}
\usepackage{alltt}
\usepackage{boxedminipage}
\usepackage{tikz}
\usepackage{calc}
\usepackage{subfloat}
\usepackage{multirow}
\usepackage{cmll}
\usepackage{multicol}
\definecolor{color1}{RGB}{204,0,51}
\definecolor{color2}{RGB}{159,182,205}
\usepackage{bookmark}

\usepackage{verbatim}
\usepackage[vcentermath]{youngtab}
\usepackage{young}
\usepackage{transparent}
\usepackage{color,soul}
\usepackage{cancel}
\usepackage{bm}
\usepackage{pifont}
\usepackage{placeins}
\usepackage{mathtools}
\usepackage{rotating}
\usepackage[refpage]{nomencl}
\usepackage{hhline}
\usepackage{marginnote}
\usepackage{upgreek}
\usepackage{stackengine}
\usepackage{ytableau}
\usepackage{listings}
\usepackage{calligra}
\usepackage{afterpage}
\usepackage{cite}
\usepackage{cleveref}
\usepackage{makecell}
\usepackage{accents}
\usepackage{breqn}
\usepackage{environ}

\usepackage{tikz-cd}
\tikzset{
    marrow/.style={decoration={markings,mark=at position 0.5 with {\arrow{#1}}}, postaction=decorate}
}

\usepackage[final]{showlabels}

\def\rme{\mathrm{e}}
\def\rmd{\mathrm{d}}

\usepackage{comment}

\usepackage{caption}
\usepackage{subcaption}
\usetikzlibrary{arrows}

\usepackage{xcolor}
\usetikzlibrary{calc}

\newcommand{\rectcol}[1]{\node[rectangle, draw={#1},very thick,minimum height=1cm,minimum width=2cm]}
\newcommand{\rectcolt}[1]{\node[rectangle, draw=black,fill={#1},minimum height=1cm,minimum width=1cm]}
\newcommand{\rectline}{\node[rectangle, draw=black, minimum height=1cm,minimum width=4cm]}
\newcommand{\rectlinea}{\node[rectangle, draw=black, minimum height=1cm,minimum width=6cm]}

\newcommand{\rectfour}[6]{
\rectcolt{{#3}} at ($(0,0)+({#1},{#2})$) {};
\rectcolt{{#4}} at ($(1,0)+({#1},{#2})$) {};
\rectcolt{{#5}} at ($(2,0)+({#1},{#2})$) {};
\rectcolt{{#6}} at ($(3,0)+({#1},{#2})$) {};
\rectline at ($(1.5,0)+({#1},{#2})$) {};
 }

\newcommand{\rectsix}[8]{
\rectcolt{{#3}} at ($(0,0)+({#1},{#2})$) {};
\rectcolt{{#4}} at ($(1,0)+({#1},{#2})$) {};
\rectcolt{{#5}} at ($(2,0)+({#1},{#2})$) {};
\rectcolt{{#6}} at ($(3,0)+({#1},{#2})$) {};
\rectcolt{{#7}} at ($(4,0)+({#1},{#2})$) {};
\rectcolt{{#8}} at ($(5,0)+({#1},{#2})$) {};
\rectlinea at ($(2.5,0)+({#1},{#2})$) {};
 }

\definecolor{darkergreen}{rgb}{0.0, 0.5, 0.0}

\allowdisplaybreaks

\sethlcolor{yellow}
\definecolor{boh}{RGB}{79,47,79}

\hypersetup{
        linkbordercolor={red}
}

\hypersetup{
unicode=false,          
pdftoolbar=true,        
pdfmenubar=true,        
pdffitwindow=false,     
pdfstartview={FitH},    
pdftitle={Generalized Newton-Cartan Geometries for Particles and Strings},    
pdfauthor={Eric Bergshoeff, Kevin van Helden, Johannes Lahnsteiner, Luca Romano, Jan Rosseel},     
pdfsubject={},   
pdfnewwindow=true,      
colorlinks=false,       
linkcolor=red,          
citecolor=red,        
filecolor=red,      
urlcolor=red           
linkbordercolor={red},
citebordercolor={red},
urlbordercolor={red}
}

\makeatletter

\newcommand{\Rmnum}[1]{\expandafter\@slowromancap\romannumeral #1@}
\makeatother

\usepackage{alphalph}

\makeatletter
\newalphalph{\aalphalph}[mult]{\alphalph@alph}{26}
\newcommand{\alphalphval}[1]{%
\@ifundefined{c@#1}{
\aalphalph{#1}
}{%
\aalphalph{\value{#1}}
}
}
\makeatother

\usepackage{color}

\def\chapterautorefname~#1\null{Chap.~(#1)\null}
\def\sectionautorefname~#1\null{Sec.~(#1)\null}
\def\subsectionautorefname~#1\null{sub--Sec.~(#1)\null}
\def\figureautorefname~#1\null{Fig.~(#1)\null}
\def\tableautorefname~#1\null{Tab.~(#1)\null}
\def\equationautorefname~#1\null{eq.~(#1)\null}

\def\equationautorefname~#1\null{eq.~(#1)\null}

\NewEnviron{multieq}[1][2]{

\begin{multicols}{#1}
\begin{subequations}
\setlength{\abovedisplayskip}{-15pt}
\allowdisplaybreaks
\begin{align}
\BODY
\end{align}
\end{subequations}
\end{multicols}
\setlength{\parindent}{0pt}
}

\DeclareMathAlphabet\mathbfcal{OMS}{cmsy}{b}{n}

\usepackage[yyyymmdd]{datetime}

\title{
\bf Generalized Newton-Cartan Geometries for Particles and Strings}
\date{}

\begin{document}

\begin{flushright}
\small
July 1\textsuperscript{st}, 2022\\
\normalsize
\end{flushright}
{\let\newpage\relax\maketitle}
\maketitle
\def\equationautorefname~#1\null{eq.~(#1)\null}
\def\tableautorefname~#1\null{tab.~(#1)\null}

\vspace{0.8cm}

\begin{center}
\renewcommand{\thefootnote}{\alph{footnote}}
{\sl\large E.~A.~Bergshoeff$^{~1}$}\footnote{Email: {\tt e.a.bergshoeff[at]rug.nl}},
{\sl\large K.~van~Helden$^{~1,2}$}\footnote{Email: {\tt k.s.van.helden[at]rug.nl}}
{\sl\large J.~Lahnsteiner$^{~1}$}\footnote{Email: {\tt j.m.lahnsteiner[at]outlook.com}},
{\sl\large L.~Romano$^{~1,3}$}\footnote{Email: {\tt lucaromano2607[at]gmail.com}},\\[.1truecm]
{\sl\large J.~Rosseel$^{~4,5}$}\footnote{Email: {\tt jan.rosseel[at]univie.ac.at}}
\setcounter{footnote}{0}
\renewcommand{\thefootnote}{\arabic{footnote}}

\vspace{0.5cm}
${}^1${\it Van Swinderen Institute, University of Groningen\\
Nijenborgh 4, 9747 AG Groningen, The Netherlands}\\
\vskip .2truecm
${}^2${\it Bernoulli Institute, University of Groningen\\
Nijenborgh 9, 9747 AG Groningen, The Netherlands}\\
\vskip .2truecm
${}^3${\it Departamento de Electromagnetismo y Electrónica,Universidad de Murcia,\\
Campus de Espinardo, 30100 Murcia, Spain}\\
\vskip .2truecm
${}^4${\it Faculty of Physics, University of Vienna,\\
Boltzmanngasse 5, 1090 Wien, Austria }\\
\vskip .2truecm
${}^5${\it Division of Theoretical Physics, Rudjer Bo\v{s}kovi\'c Institute,\\
Bijeni\v{c}ka 54, 10000 Zagreb, Croatia}\\

\vspace{1.8cm}


{\bf Abstract}
\end{center}
\begin{quotation}
{\small \noindent
We discuss the generalized Newton-Cartan geometries that can serve as gravitational background fields for particles and strings. In order to enable us to define affine connections that are invariant under all the symmetries of the structure group, we describe torsionful geometries with independent torsion tensors. A characteristic feature of the non-Lorentzian geometries we consider is that some of the torsion tensors are so-called `intrinsic torsion' tensors that cannot be absorbed in any of the spin connections. Setting some components of these intrinsic torsion tensors to zero leads to constraints on the geometry. For both particles and strings, we discuss various such constraints that can be imposed consistently with the structure group symmetries. In this way, we reproduce several results in the literature.
}
\end{quotation}

\newpage

\tableofcontents

\section{Introduction}\label{sec:intro}

\noindent One of the cornerstones of Einstein's description of general relativity is its underlying semi-Riemannian geometry giving a geometrical interpretation to the gravitational force. What is less known is that also Newtonian gravity can be given a geometrical interpretation using a degenerate foliated geometry. Its proper formulation was given eight years after Einstein's formulation by \'Elie Cartan \cite{Cartan1,Cartan2}. This generalization of Newtonian gravity is valid in any frame, includes strong gravitational effects and is called Newton-Cartan (NC) gravity with an underlying geometry that is called NC geometry. This is the correct geometry to describe the coupling of gravity to massive non-relativistic particles and field theories.

Recently, there has been a growing interest in other non-Lorentzian\,\footnote{We will generically call any gravity theory with a structure group that differs from the Lorentz group `non-Lorentzian'. However, for historic reasons we will instead sometimes use the denomination `non-relativistic' for NC gravity and its generalization to strings. In this context we will also use the wording `non-relativistic string theory' as a candidate theory of `non-relativistic quantum gravity'.} gravity models and corresponding geometries. One key example is an extension of Newtonian gravity including so-called `twistless torsion' that was shown to occur in Lifshitz holography where it was realized as a background geometry of the boundary conformal field theory \cite{Christensen:2013lma}. This is a natural extension since the twistless torsion condition is invariant under (anisotropic) local dilatations, as it should for a Lifshitz conformal field theory, whereas the zero torsion condition describing a Newtonian space-time is not. For a useful review on general non-Lorentzian holography, see \cite{Zaanen:2015oix}. Another interesting non-Lorentzian geometry is Carroll geometry, which appears as the natural geometry of null surfaces \cite{Duval:2014uoa}; see also the recent paper \cite{Figueroa-OFarrill:2022mcy} and references therein.

Another way to generalize NC geometry is to go beyond particles and consider the gravitational coupling to extended objects such as strings. Whereas any extended object can be coupled to general relativity, in the non-Lorentzian case each extended object requires a different non-Lorentzian geometry with a foliation that is determined by the spatial extension of the object: particles require a foliation with leaves of co-dimension one, but strings require a foliated geometry where the leaves are submanifolds of codimension two, that describe the dimensions transversal to the string. This geometry is not only relevant to describe the coupling of non-Lorentzian gravity to a classical cosmic string but can also be used to formulate the sigma model describing non-relativistic string theory in a general curved background.\,\footnote{By a non-relativistic string we mean a string with a relativistic worldsheet that propagates in a non-relativistic target space.} Originally, non-relativistic string theory was only formulated for a flat non-Lorentzian space-time \cite{Gomis:2000bd,Danielsson:2000gi} or special backgrounds \cite{Gomis:2005pg}.  Only recently a formulation for a generic background has been given \cite{Bergshoeff:2018yvt,Bergshoeff:2019pij}. This opens the way to study essential features of non-relativistic string theory as a candidate theory of non-relativistic quantum gravity, independent of the relativistic case. The geometry underlying non-relativistic string theory has natural torsion tensors that are constrained by requiring that the quantum effective action remains non-relativistic \cite{Yan:2021lbe} and/or requiring supersymmetry \cite{Bergshoeff:2021tfn}. For recent reviews on non-relativistic string theory and non-Lorentzian geometries with more references, see \cite{Oling:2022fft, Bergshoeff:2022eog}.

An important feature of NC gravity and its generalization to strings is that its coupling to particles and/or strings is described by additional fields beyond the usual frame fields. In the case of particles this extra field is a 1-form, called $m_\mu$. This 1-form field  is needed due to the fact that, unlike in the relativistic case, mass and energy are two distinct conserved quantities in the non-relativistic case. It has a clear algebraic interpretation as the gauge field associated with the central extension that distinguishes the Galilei from the Bargmann algebra. It couples to a particle via a Wess-Zumino term. In the case of (bosonic) string theory these extra fields are the non-relativistic Kalb-Ramond 2-form $b_{\mu\nu}$ and the dilaton $\phi$. Like in the particle case, the 2-form $b_{\mu\nu}$ couples to a non-relativistic string via a Wess-Zumino term. Both $m_\mu$ and $b_{\mu\nu}$ have in common that they are part of the geometric fields in the sense that they vary under boost transformations and, in fact, are needed to write down boost-invariant actions describing the coupling to particles and/or strings.

When discussing torsionful geometries it is important to distinguish between the relativistic and non-Lorentzian case. In the relativistic case a torsion tensor can always be viewed as a deformation of an affine connection that can be removed without imposing any constraints on the metric structure. This is no longer the case in non-Lorentzian geometry. There, part of the torsion consists of so-called intrinsic torsion tensor components that form an obstruction to defining a metric compatible and torsionless connection, without imposing differential constraints on the metric structure \cite{Figueroa-OFarrill:2020gpr}. In the physics literature, these intrinsic torsion tensors are sometimes introduced as dependent tensors that are expressed in terms of other (e.g. geometric) fields of the model \cite{Hartong:2015zia}. It is the purpose of this work to introduce torsion tensors as independent fields in the spirit of \cite{Geracie:2015dea, Bekaert:2014bwa,Bekaert:2015xua,Pereniguez:2019eoq} and demonstrate which of these tensors can be set to zero consistently with the symmetries of the structure group. A benefit of this approach is that in this way we are always able to define a proper affine connection that is invariant under all the symmetries of the structure group. Only afterwards we will try to express some of the torsion tensors as dependent tensors in terms of the other (geometric) fields of the model. Both for particles and strings we will give explicit examples of such dependent intrinsic torsion tensors.

This work is organized as follows. In section 2 we review, using the Cartan frame formulation, the torsionful NC geometry appropriate for describing the coupling of gravity to non-relativistic massive particles. In particular, we show how to introduce independent torsion tensors and how these tensors can be used to define spin connections and an invariant affine connection. We discuss various constraints that can be imposed on these torsion tensors without breaking the symmetries of the structure group. We end this section by giving several examples of dependent torsion tensors that have appeared in the literature. In section 3 we extend all calculations of section 2 from particles to strings. Apart from getting slightly more involved the structure of this section is very similar to that of section 2. Finally, in section 4 we give our conclusions and present an outlook for future extensions.

\section{Torsionful Newton-Cartan geometry in the Cartan formulation} \label{sec:TNC}

\noindent Newton-Cartan (NC) geometry refers to the geometry of $D$-dimensional manifolds, called NC manifolds, that are equipped with a degenerate metric structure that reduces the local structure group to the homogeneous Galilei group in $D$ dimensions. The latter is given by the semi-direct product SO$(D-1) \rtimes \mathbb{R}^{D-1}$, where SO$(D-1)$ is physically interpreted as the group of local spatial rotations and $\mathbb{R}^{D-1}$ as that of local Galilean boosts. A convenient way to introduce the metric and metric compatible affine connection structures of NC geometry is in terms of the frame and structure group connection fields of a Cartan formulation.

The Cartan formulation of NC geometry, for both torsionless and torsionful affine connections, has been developed in a number of references \cite{Duval:1984cj,Andringa:2010it,Christensen:2013lma,Christensen:2013rfa,Hartong:2014pma,Bergshoeff:2014uea,Geracie:2015dea,Hartong:2015zia}. Here, we will review it to facilitate generalization to the case of String Newton-Cartan (SNC) geometry. We will first introduce the frame fields and ensuing metric structure in section \ref{ssec:framemetricTNC}. In section \ref{ssec:connectionTNC}, we will discuss how torsionful, metric compatible affine connections, that are completely determined in terms of the frame fields and suitable torsion tensors, can be defined on a NC manifold. We will at first treat the torsion in a general manner, by introducing it as an extra, independent ingredient. Special cases and examples of torsionful NC geometry that appear in the context of e.g. Lifshitz holography \cite{Christensen:2013lma,Christensen:2013rfa,Hartong:2014pma,Bergshoeff:2014uea,Hartong:2015zia}, will be discussed in section \ref{ssec:specialTNC}.

\subsection{Frame fields and metric structure} \label{ssec:framemetricTNC}

\noindent The frame fields on a $D$-dimensional NC manifold $\mathcal{M}$, with local coordinates $x^\mu$, consist of a so-called `time-like Vielbein' or `clock form' $\tau_\mu$ and a `spatial Vielbein' $e_\mu{}^a$ ($a = 1, \cdots, D-1$).\footnote{Technically speaking, the Vielbeine are sections of the coframe bundle and not the frame bundle. In this work, we follow the physics literature where these fields are often called frame fields.} In addition to that, we introduce a so-called `mass form' $m_\mu$ as part of the geometric data.\footnote{In approaches to define NC geometry as a gauging of the Bargmann algebra, i.e., the centrally extended Galilei algebra, the mass form $m_\mu$ corresponds to the gauge field associated with the central extension \cite{Duval:1984cj,Andringa:2010it} For this reason, $m_\mu$ is often called the `central charge gauge field' in the literature.}. In what follows, the spatial index $a$ will be freely raised and lowered using Kronecker deltas $\delta_{ab}$ and $\delta^{ab}$. The one-forms $\tau_\mu$, $e_\mu{}^a$ and $m_\mu$ transform in a reducible, indecomposable representation of the structure group, according to the following infinitesimal local transformation rules:
\begin{align}
  \label{eq:localtrafosindep}
  \delta \tau_\mu &= 0 \,, \qquad \qquad \qquad
  \delta e_\mu{}^a = - \lambda^{ab} e_{\mu b} + \lambda^a \tau_\mu \,, \qquad \qquad \qquad
  \delta m_\mu = \lambda^a e_{\mu a} \,.
\end{align}
Here, $\lambda^{ab} = -\lambda^{ba}$ denote the parameters of SO$(D-1)$ spatial rotations, whereas $\lambda^a$ refer to the parameters of $\mathbb{R}^{D-1}$ Galilean boosts. In order to define the metric structure on a NC geometry, it is convenient to also introduce vector fields $\tau^\mu$, $e_a{}^\mu$ via the following relations:
\begin{alignat}{3}
  \label{eq:invVielbeine}
  & \tau^\mu \tau_\mu = 1\,, \qquad \qquad \qquad & & \tau^\mu e_\mu{}^a = 0 \,, \qquad \qquad \qquad \qquad & & e_a{}^\mu \tau_\mu = 0 \,, \nonumber \\
  & e_\mu{}^a e_b{}^\mu = \delta_b^a \,, \qquad \qquad \qquad & & \tau_\mu \tau^\nu + e_\mu{}^a e_a{}^\nu = \delta_\mu^\nu \,.
\end{alignat}
These formulas express that the square matrices $\begin{pmatrix} \tau_\mu & e_\mu{}^a\end{pmatrix}$ and $\begin{pmatrix} \tau^\mu \\ e_a{}^\mu\end{pmatrix}$ are each other's inverse and we will henceforth (with slight abuse of terminology) refer to $\tau^\mu$ as the `inverse time-like Vielbein' and to $e_a{}^\mu$ as the `inverse spatial Vielbein'.
Their transformation rules under local Galilean boosts and spatial rotations are given by:
\begin{align}
  \label{eq:localtrafosinv}
  \delta \tau^\mu &= -\lambda^a e_a{}^\mu \,, \qquad \qquad \qquad \delta e_a{}^\mu = - \lambda_a{}^b e_b{}^\mu \,.
\end{align}
One can then construct two degenerate symmetric (covariant and contravariant) two-tensors that are invariant under local rotations and boosts:
\begin{align} \label{eq:TNCmetricstruct}
  \tau_{\mu\nu} \equiv \tau_\mu \tau_\nu \,, \qquad \qquad \qquad  h^{\mu\nu} \equiv e_a{}^{\mu} e_b{}^\nu \delta^{ab} \,.
\end{align}
These define the degenerate metric structure on the NC manifold $\mathcal{M}$. The covariant metric $\tau_{\mu\nu}$ has rank 1 and is referred to as the `time-like metric', whereas the contravariant metric $h^{\mu\nu}$ has rank $D-1$ and is often called the `spatial co-metric'. Note that $\tau_\mu$ is in the kernel of the spatial co-metric, i.e., $h^{\mu\nu} \tau_\nu = 0$, as a consequence of \eqref{eq:invVielbeine}.

The local causal structure of a NC manifold can be viewed as a limit of that of a Lorentzian manifold, in which the speed of light in a local inertial reference frame is sent to infinity. In local Minkowskian coordinates $\{x^0, x^a\}$, this can be achieved by rescaling $x^0$ with a (dimensionless) parameter $\omega$ and taking the limit $\omega \rightarrow \infty$. In this limit the local lightcone $\omega^2 (x^0)^2  = x^a x_a$ flattens out and degenerates into the $x^0=0$ hyperplane. With respect to such a local inertial reference frame, vectors can be classified as time-like future-/past-directed, when they have a strictly positive/negative $x^0$-component and as spatial when their $x^0$-component is zero. This can be phrased covariantly, using the time-like Vielbein $\tau_\mu$, by saying that a vector $X^\mu$ is time-like future-directed when $\tau_\mu X^\mu > 0$, time-like past-directed when $\tau_\mu X^\mu <0$ and spatial when $\tau_\mu X^\mu = 0$. Here, it is understood that $X^\mu$ itself is invariant under the structure group.

As their names suggest, the symmetric two-tensors $\tau_{\mu\nu}$ and $h^{\mu\nu}$ allow one to compute time intervals and spatial distances in NC geometry in a way that is analogous to how the metric is used to calculate lengths of curves in Riemannian geometry \cite{Dombrowski:1964,Kunzle:1972}. Time intervals in the NC manifold $\mathcal{M}$ are defined along any curve $\gamma : t \in [0,1] \rightarrow x^\mu(t) \in \mathcal{M}$, whose tangent vectors $\dot{x}^\mu(t) \equiv \rmd x^\mu(t)/\rmd t$ are time-like future-directed for all $t \in (0,1)$. Such a curve models the motion of a non-relativstic physical particle or observer between two points with local coordinates $x^\mu(0)$ and $x^\mu(1)$. The time interval needed by the particle/observer to traverse the curve $\gamma$ is then computed using the time-like metric $\tau_{\mu\nu}$
\begin{align}
  \label{eq:timeinterval}
  \int_0^1 \rmd t \, \sqrt{\dot{x}^\mu \dot{x}^\nu \tau_{\mu\nu}} = \int_0^1 \rmd t \, \dot{x}^\mu \tau_\mu = \int_\gamma \rmd x^\mu \tau_\mu \,.
\end{align}

To define spatial distances in an analogous way, one needs an inverse of the spatial co-metric $h^{\mu\nu}$. The latter is not invertible when viewed as a map from one-forms to vectors, since it has a non-trivial kernel spanned by $\tau_\mu$. It does however give rise to a well-defined map between the space of equivalence classes $[\alpha_\mu] = \{\alpha_\mu + f \tau_\mu | f \in C^\infty(\mathcal{M})\}$ of one-forms that differ by a multiple of $\tau_\mu$, i.e., $f\tau_\mu$, and the space of spatial vectors, where $h^{\mu\nu}$ maps $[\alpha_\nu]$ to $h^{\mu\nu}[\alpha_\nu] \equiv h^{\mu\nu} \alpha_\nu$. When viewed like this, $h^{\mu\nu}$ is invertible and its inverse is given by
\begin{align}
  h_{\mu\nu} = e_\mu{}^a e_\nu{}^b \delta_{ab} \,.
\end{align}
Here, $h_{\mu\nu}$ is interpreted as a map that assigns the equivalence class $[h_{\mu\nu} X^\nu]$ to each spatial vector $X^\mu$. Using \eqref{eq:invVielbeine}, one sees that
\begin{align}
  h^{\mu\rho} h_{\rho\nu} = \delta^\mu_\nu - \tau^\mu \tau_\nu \,,
\end{align}
from which it follows that $h^{\mu\rho} h_{\rho\nu}$ and $h_{\nu\rho} h^{\rho\mu}$ act as the identity $\delta^\mu_\nu$ on spatial vectors $X^\nu$, resp. equivalence classes $[\alpha_\mu]$. The two-tensors $h^{\mu\nu}$ and $h_{\mu\nu}$ are thus indeed each other's inverse, when viewed as maps between the space of equivalence classes of one-forms that are equal up to a multiple of $\tau_\mu$ and the space of spatial vectors. It is also worth mentioning that $h_{\mu\nu}$ (unlike $h^{\mu\nu}$) is not invariant under local boosts: $\delta h_{\mu\nu} = 2 \lambda^a \tau_{(\mu} e_{\nu) a}$. It thus does not give a covariant metric on the full space of vectors. Note however that $X^\mu Y^\nu h_{\mu\nu}$ is boost invariant when $X^\mu$ and $Y^\nu$ are spatial\footnote{A slightly stronger statement is that the equivalence class $[h_{\mu\nu} X^\nu]$ is boost invariant, when $X^\mu$ is spatial.}, so that $h_{\mu\nu}$ constitutes a covariant metric (with Euclidean signature) on the space of spatial vectors. With these remarks in mind, spatial distances can be defined along any curve $\tilde{\gamma} : s \in [0,1] \rightarrow x^\mu(s)$, whose tangent vectors $x^{\prime \mu}(s) \equiv \rmd x^\mu(s)/\rmd s$ are spatial for all $s \in (0,1)$. The length of such a curve is defined in terms of the metric $h_{\mu\nu}$ on spatial vectors as:
\begin{align}
  \label{eq:length}
  \int_0^1 \rmd s \, \sqrt{x^{\prime \mu} x^{\prime \nu} h_{\mu\nu}} \,.
\end{align}
The notion that one can only measure lengths between simultaneous events is formalized by the fact that spatial distances can only be defined along curves whose tangent vectors are spatial.

The above discussion shows that the frame fields $\tau_\mu$ and $e_\mu{}^a$ can be viewed as a non-relativistic analogue of the Vielbein of Lorentzian geometry in the Cartan formulation. The frame field $m_\mu$ has no analogue in Lorentzian geometry. It is not needed to specify the metric structure of NC geometry. It however plays an important role in defining metric compatible affine connections that are fully expressed in terms of frame fields and torsion tensors, as we will review in the next section.

\subsection{Torsionful, metric compatible connection} \label{ssec:connectionTNC}

\noindent In the Cartan formulation of NC geometry, metric compatible affine connections are defined by introducing a structure group connection one-form $\Omega_\mu$ that takes values in the Lie algebra of the homogeneous Galilei group in $D$ dimensions:
\begin{equation}
  \Omega_\mu = \frac12 \omega_\mu{}^{ab} J_{ab} + \omega_\mu{}^a G_a \,,
\end{equation}
where $J_{ab} = -J_{ba}$ and $G_a$ are generators of the Lie algebra of SO$(D-1)$ (spatial rotations) and $\mathbb{R}^{D-1}$ (Galilean boosts). We will refer to $\omega_\mu{}^{ab} = - \omega_\mu{}^{ba}$ and $\omega_\mu{}^a$ as the spin connections for spatial rotations and Galilean boosts respectively. Their infinitesimal local structure group transformations are given by
\begin{align}
  \label{eq:localtrafosdep}
    \delta \omega_\mu{}^{ab} &= \partial_\mu \lambda^{ab} - 2 \lambda^{[a|c|} \omega_{\mu c}{}^{b]} \,, \qquad \qquad \qquad
  \delta \omega_\mu{}^a = \partial_\mu \lambda^a + \omega_\mu{}^{ab} \lambda_b - \lambda^{ab} \omega_{\mu b} \,.
\end{align}

To introduce an affine connection $\Gamma^\rho_{\mu\nu}$ that is compatible with the NC metric structure, one then considers the following `Vielbein postulates':
\begin{align}
  \label{eq:Vielbpost}
  & \partial_\mu \tau_\nu - \Gamma_{\mu\nu}^\rho \tau_\rho = 0 \,, \qquad \qquad \qquad \qquad
   \partial_\mu e_\nu{}^a + \omega_\mu{}^{ab} e_{\nu b} - \omega_\mu{}^a \tau_\nu - \Gamma_{\mu\nu}^\rho e_{\rho}{}^a = 0 \,,
\end{align}
from which metric compatibility
\begin{align}
  \label{eq:metricscomp}
  \nabla_\mu \tau_{\nu\rho } & \equiv \partial_\mu \tau_{\nu\rho} - \Gamma_{\mu\nu}^\sigma \tau_{\sigma \rho} - \Gamma_{\mu\rho}^\sigma \tau_{\nu \sigma} = 0 \,, \qquad \qquad \nabla_\mu h^{\nu\rho}  \equiv \partial_\mu h^{\nu\rho} + \Gamma_{\mu\sigma}^\nu h^{\sigma \rho} + \Gamma_{\mu\sigma}^\rho h^{\nu\sigma}= 0 \,.
\end{align}
immediately follows. The connection $\Gamma^\rho_{\mu\nu}$ is taken to have the appropriate transformation law under general coordinate transformations and to be invariant under local spatial rotations and Galilean boosts. The set of Vielbein postulates \eqref{eq:Vielbpost} is then invariant under these transformations; in particular, one finds that the second postulate transforms to the first under boosts. One can use \eqref{eq:Vielbpost} to express $\Gamma^\rho_{\mu\nu}$ in terms of the spin connections $\omega_\mu{}^{ab}$, $\omega_\mu{}^a$ and the time-like and spatial Vielbeine $\tau_\mu$, $e_\mu{}^a$ as follows:
\begin{align} \label{eq:GammaVielbpost}
  \Gamma^\rho_{\mu\nu} = \tau^\rho \partial_\mu \tau_\nu + e_a{}^\rho \left( \partial_\mu e_\nu{}^a + \omega_\mu{}^{ab} e_{\nu b} - \omega_\mu{}^a \tau_\nu \right) \,.
\end{align}
Note that this is invariant under the local rotation and boost transformations  \eqref{eq:localtrafosindep}, \eqref{eq:localtrafosinv} and \eqref{eq:localtrafosdep}. In other words, the affine connection $\Gamma^\rho_{\mu\nu}$ is invariant under the local structure group.

So far, we have not imposed any restrictions on the torsion $2 \Gamma_{[\mu\nu]}^\rho$ of the affine connection $\Gamma^\rho_{\mu\nu}$. In this section, we will keep the torsion completely arbitrary and view it as an extra independent geometric ingredient. It is then convenient to decompose it in two a priori independent tensors $T_{\mu\nu}$ and $T_{\mu\nu}{}^a$ as follows:
\begin{align} \label{eq:torsiondecomp}
  & 2 \Gamma^\rho_{[\mu\nu]} = \tau^\rho T_{\mu\nu} + e_a{}^\rho T_{\mu\nu}{}^a \qquad \Leftrightarrow \qquad
   T_{\mu\nu}  \equiv 2 \Gamma^\rho_{[\mu\nu]} \tau_\rho  \quad \text{and} \quad T_{\mu\nu}{}^a = 2 \Gamma^\rho_{[\mu\nu]} e_\rho{}^a \,.
\end{align}
We will refer to $T_{\mu\nu}$ and $T_{\mu\nu}{}^a$ as the time-like and spatial torsion respectively. They transform under local rotations and boosts as:
\begin{align}
  \label{eq:trafoTTa}
  \delta T_{\mu\nu} = 0 \,, \qquad \qquad \qquad \delta T_{\mu\nu}{}^a = - \lambda^a{}_b T_{\mu\nu}{}^b + \lambda^a T_{\mu\nu} \,.
\end{align}

By antisymmetrizing the Vielbein postulates \eqref{eq:Vielbpost}, one obtains the following equations that are covariant with respect to local spatial rotations and Galilean boosts:
\begin{align}
  \label{eq:torsionVielb}
  2 \partial_{[\mu} \tau_{\nu]} = T_{\mu\nu}\,, \qquad \qquad \qquad 2 \partial_{[\mu} e_{\nu]}{}^a + 2 \omega_{[\mu}{}^{ab} e_{\nu] b} - 2 \omega_{[\mu}{}^a \tau_{\nu]} = T_{\mu\nu}{}^a \,.
\end{align}
The second of these should be viewed as an identity that exhibits that some components of $\omega_\mu{}^{ab}$ and $\omega_\mu{}^a$ are not independent fields. Viewing it as a system of algebraic equations for $\omega_\mu{}^{ab}$ and $\omega_\mu{}^a$, one can solve it to express some of the spin connection components in terms of $\tau_\mu$, $e_\mu{}^a$ and the torsion tensor $T_{\mu\nu}{}^a$. The spatial torsion $T_{\mu\nu}{}^a$ is thus absorbed in these expressions for the spin connections. This is analogous to what happens in torsionful Lorentzian geometry, where the spin connection for Lorentz transformations can be determined in terms of the Vielbein and the torsion. In the Lorentzian case, the torsion is fully absorbed in the spin connection expression. This is no longer the case in NC geometry, as the first equation of \eqref{eq:torsionVielb} shows: the time-like torsion $T_{\mu\nu}$ cannot be absorbed in any of the spin connections $\omega_\mu{}^{ab}$, $\omega_\mu{}^a$. For this reason, it is also called the `intrinsic torsion' of NC geometry \cite{Figueroa-OFarrill:2020gpr}. Unlike the second equation, that reduces to an identity when (certain components of) $\omega_\mu{}^{ab}$ and $\omega_\mu{}^a$ are expressed in terms of $\tau_\mu$, $e_\mu{}^a$ and $T_{\mu\nu}{}^a$, the first equation of \eqref{eq:torsionVielb} represents a geometric constraint that equates the curl of the time-like Vielbein to the time-like torsion. Such a constraint has no analogue in Lorentzian geometry.

In Lorentzian geometry, the metric compatible spin and affine connections are completely determined by the Vielbeine or metric and the torsion. By contrast, in NC geometry it is not possible to specify the spin and affine connections, introduced above, solely in terms of the torsion and time-like and spatial Vielbeine/metrics. While we already noted that the second equation of \eqref{eq:torsionVielb} can be solved to express some components of $\omega_\mu{}^{ab}$ and $\omega_\mu{}^a$ as functionals of $\tau_\mu$, $e_\mu{}^a$ and $T_{\mu\nu}{}^a$, it cannot be used to give all spin connection components in this way. Indeed, the fields $\omega_\mu{}^{ab}$ and $\omega_\mu{}^a$ contain $D(D-1)(D-2)/2 + D(D-1) = D^2(D-1)/2$ components in total, while the second equation of \eqref{eq:torsionVielb} only has $D(D-1)^2/2$ components. Consequently, $D(D-1)/2$ spin connection components cannot be solved in terms of Vielbeine and $T_{\mu\nu}{}^a$ from this equation. It is however possible to express these remaining components as function of torsion tensors, Vielbeine, and the mass form $m_\mu$. To do this, we introduce an extra `mass torsion tensor' $T^{(m)}_{\mu\nu}$ \footnote{We use the term `torsion tensor' for $T^{(m)}_{\mu\nu}$ in the Cartan formulation sense, namely as corresponding to the properly covariantized curl of a geometric field. It should be emphasized that $T^{(m)}_{\mu\nu}$ does not correspond to components of the torsion of the affine connection $\Gamma^\rho_{\mu\nu}$. The latter is fully captured by $T_{\mu\nu}$ and $T_{\mu\nu}{}^a$. As will be seen in section \ref{ssec:specialTNC}, the introduction of $T^{(m)}_{\mu\nu}$ is necessary to consistently include unconstrained torsion of $\Gamma^\rho_{\mu\nu}$.} and equate it to the properly covariantized (with respect to Galilean boosts) curl of the mass form field $m_\mu$:
\begin{align}
  \label{eq:extraconvconstraint}
  2 \partial_{[\mu} m_{\nu]} - 2 \omega_{[\mu}{}^a e_{\nu] a} = T^{(m)}_{\mu\nu} \,.
\end{align}
The mass torsion tensor is invariant under spatial rotations and transforms under Galilean boosts as follows:
\begin{align} \label{eq:trafoTM}
  \delta T^{(m)}_{\mu\nu} = \lambda_a T_{\mu\nu}{}^a \,.
\end{align}
Note that the left-hand side of \eqref{eq:extraconvconstraint} is invariant under an abelian gauge transformation (with parameter $\sigma$) of $m_\mu$:
\begin{align}
  \label{eq:centralcharge}
  \delta m_\mu = \partial_\mu \sigma \,.
\end{align}
This transformation is often called the `central charge transformation'.\footnote{The terminology stems from the approach in which Newton-Cartan geometry is defined via a gauging of the centrally extended Galilei algebra. The transformation \eqref{eq:centralcharge} then corresponds to the central extension; see also footnote 1.}

The two equations
\begin{align}
  \label{eq:conventional}
  2 \partial_{[\mu} e_{\nu]}{}^a + 2 \omega_{[\mu}{}^{ab} e_{\nu] b} - 2 \omega_{[\mu}{}^a \tau_{\nu]} = T_{\mu\nu}{}^a\,, \qquad \qquad \qquad 2 \partial_{[\mu} m_{\nu]} - 2 \omega_{[\mu}{}^a e_{\nu] a} = T^{(m)}_{\mu\nu}
\end{align}
then constitute a system of $D(D-1)^2/2 + D(D-1)/2 = D^2(D-1)/2$ algebraic equations for as many components of $\omega_\mu{}^{ab}$ and $\omega_\mu{}^a$. Solving these equations then leads to the following expressions for the spin connections in terms of the frame fields and torsion tensors $T_{\mu\nu}{}^a$, $T^{(m)}_{\mu\nu}$:
\begin{align}
  \label{eq:spinconnexpr}
  \omega_\mu{}^{a} &= \tau_\mu \tau^\nu e^{a \rho} \partial_{[\nu} m_{\rho]} + e^{a \nu} \partial_{[\mu} m_{\nu]} + e_{\mu b} e^{a \nu} \tau^\rho \partial_{[\nu} e_{\rho]}{}^b + \tau^\nu \partial_{[\mu} e_{\nu]}{}^a \nonumber \\ & \qquad - \tau_\mu \tau^\nu e^{a\rho} T^{(m)}_{\nu\rho} + e_{\mu b} \tau^\nu e^{(a|\rho|} T_{\nu\rho}{}^{b)} - \frac12 e_{\mu b} e^{b \nu} e^{a\rho} T^{(m)}_{\nu\rho} \,, \nonumber \\
  \omega_\mu{}^{ab} &= 2 e^{[a|\nu|} \partial_{[\mu} e_{\nu]}{}^{b]} - e_{\mu c} e^{a\nu} e^{b\rho} \partial_{[\nu} e_{\rho]}{}^c + \tau_\mu e^{a\nu} e^{b\rho} \partial_{[\nu} m_{\rho]} \nonumber \\ & \qquad - \frac12 \tau_\mu e^{a\nu} e^{b\rho} T^{(m)}_{\nu\rho} - e^{[a|\nu|}  T_{\mu\nu}{}^{b]} + \frac12 e_{\mu c} e^{a\nu} e^{b\rho} T_{\nu\rho}{}^c \,.
\end{align}
Plugging these expressions in \eqref{eq:GammaVielbpost}, one can express $\Gamma_{\mu\nu}^\rho$ in terms of the NC metric structure, $m_\mu$ and torsion tensors:
\begin{align}
  \label{eq:Gammaexpr}
  \Gamma_{\mu\nu}^\rho &= \tau^\rho \partial_{\mu} \tau_{\nu} + \frac12 h^{\rho \sigma} \left( \partial_\mu h_{\sigma \nu} + \partial_\nu h_{\mu\sigma} - \partial_\sigma h_{\mu\nu} \right) + h^{\rho\sigma} \tau_\mu \partial_{[\sigma} m_{\nu]} + h^{\rho\sigma} \tau_\nu \partial_{[\sigma} m_{\mu]} \nonumber \\
  & \qquad + h^{\rho\sigma} \tau_{(\mu} T^{(m)}_{\nu)\sigma} - h^{\rho\sigma} e_{(\mu|a|} T_{\nu)\sigma}{}^a + \frac12 e_a{}^\rho T_{\mu\nu}{}^a \,.
\end{align}
Taking the anti-symmetric part in $[\mu\nu]$ of this equation, one explicitly sees that the torsion $2 \Gamma^\rho_{[\mu\nu]}$ is given by \eqref{eq:torsiondecomp}, with $T_{\mu\nu}$ given by $2 \partial_{[\mu}\tau_{\nu]}$ as in \eqref{eq:torsionVielb}. The formula \eqref{eq:Gammaexpr} for $\Gamma^\rho_{\mu\nu}$ is boost invariant, although not manifestly so. One can rewrite it in a form that is manifestly invariant under local boosts as follows:
\begin{align}
  \label{eq:Gammaexpr2}
  \Gamma_{\mu\nu}^\rho &= {\bar\tau}^\rho \partial_{(\mu} \tau_{\nu)} + \frac12 h^{\rho \sigma} \left( \partial_\mu {\bar h}_{\sigma \nu} + \partial_\nu {\bar h}_{\mu\sigma} - \partial_\sigma {\bar h}_{\mu\nu} \right) + \frac12\tau^\rho T_{\mu\nu} \nonumber \\
  & \qquad + h^{\rho\sigma}m_{(\mu}T_{\nu)\sigma} + h^{\rho\sigma} \tau_{(\mu} T^{(m)}_{\nu)\sigma} - h^{\rho\sigma} e_{(\mu|a|} T_{\nu)\sigma}{}^a + \frac12 e_a{}^\rho T_{\mu\nu}{}^a \,,
\end{align}
where
\begin{equation}
{\bar \tau}^\mu = \tau^\mu + h^{\mu\nu}m_\nu\,,\hskip 2truecm {\bar h}_{\mu\nu} = h_{\mu\nu} -2 m_{(\mu}\tau_{\nu)}\,,
\end{equation}
are boost invariant expressions.

Note that eqs. \eqref{eq:conventional}, that are used to solve for the spin connections, form, together with the first of eqs. \eqref{eq:torsionVielb}, a set of equations that is invariant under the structure group. The transformation rule \eqref{eq:trafoTM} of $T^{(m)}_{\mu\nu}$ has been chosen such that this is the case. This property guarantees that the transformation rules, induced by \eqref{eq:localtrafosindep}, of the expressions \eqref{eq:spinconnexpr} for the spin connections are still given by \eqref{eq:localtrafosdep}. In the next section, we consider constraints on the torsion tensors that are consistent with the local structure group. From direct inspection or via a similar invariance argument, one also sees that the spin connection expressions \eqref{eq:spinconnexpr} are invariant under the central charge transformation \eqref{eq:centralcharge}, provided the torsion tensors $T_{\mu\nu}$, $T_{\mu\nu}{}^a$ and $T^{(m)}_{\mu\nu}$ are. The same remark holds for the affine connection expressions \eqref{eq:Gammaexpr}, \eqref{eq:Gammaexpr2}. While invariance under the central charge transformation is not manifest in \eqref{eq:Gammaexpr2}, it is manifestly realized in \eqref{eq:Gammaexpr}.

Starting from the affine connection \eqref{eq:Gammaexpr}, \eqref{eq:Gammaexpr2}, one can construct the Riemann and Ricci tensors in the usual way. The metric structure \eqref{eq:TNCmetricstruct} and affine connection \eqref{eq:Gammaexpr}, \eqref{eq:Gammaexpr2} thus fully specify torsionful NC geometry in terms of the frame fields $\tau_\mu$, $e_\mu{}^a$, $m_\mu$ and torsion tensors $T_{\mu\nu}$, $T_{\mu\nu}{}^a$, $T^{(m)}_{\mu\nu}$.

Before discussing various special cases and examples that have appeared in the recent literature, let us remark that NC geometry is the natural framework to describe the mechanics of non-relativistic point particles. A point particle traces out a worldline in space-time and, as remarked in section \ref{ssec:framemetricTNC}, time intervals along such a worldline and spatial distances to it can be measured with the metrics $\tau_{\mu\nu}$ and $h^{\mu\nu}$. The mass form $m_\mu$ also has a natural particle interpretation. Unlike relativistic theories, non-relativistic theories exhibit mass conservation. The inclusion of $m_\mu$ among the frame fields and the invariance of NC geometry under the extra central charge transformation \eqref{eq:centralcharge} then gives an extra ingredient to implement the conservation of mass of a non-relativistic particle. Given a particle with mass $m$ that moves along a worldline $\gamma : \mathbb{R}\ni t \rightarrow x^\mu(t) \in \mathcal{M}$, this can be done by introducing the following coupling to $m_\mu$
\begin{align} \label{eq:mcoupling}
   m \int_\gamma \rmd x^\mu \, m_\mu = m \int_{\mathbb{R}} \rmd t \, \dot{x}^\mu m_\mu \,.
\end{align}
This coupling of $m_\mu$ to the particle's mass current is analogous to how an electrically charged relativistic particle couples to the electromagnetic gauge potential. Gauge invariance of the coupling \eqref{eq:mcoupling} under the central charge transformation \eqref{eq:centralcharge} then implies conservation of the particle's mass current, in analogy to how charge conservation is realized in electromagnetism. We thus see that the presence of $m_\mu$ in NC geometry is natural both from the mathematical and physical point of view. Mathematically, $m_\mu$ is needed because metric compatibility no longer completely fixes the connection in terms of the metric and torsion, in case the metric structure is a degenerate non-relativistic one. Physically, it plays the role of a gauge field that couples to the Noether current that implements mass conservation.

\subsection{Special cases and examples} \label{ssec:specialTNC}

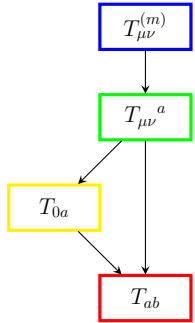
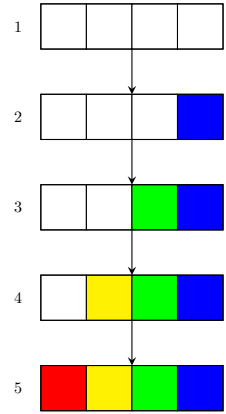
\begin{figure}[htp]
\begin{subfigure}[t]{0.45\textwidth}
\centering
\begin{tikzpicture}[scale=0.6,every node/.style={scale=0.6}]
\rectcol{blue} (blue) at (0,0) {\Large $T^{(m)}_{\mu\nu}$};
\rectcol{green} (green) at (0,-2) {\Large ${T_{\mu\nu}}^a$};
\rectcol{yellow} (yellow) at (-2,-4) {\Large $T_{0a}$};
\rectcol{red} (red) at (0,-6) {\Large $T_{ab}$};

\draw[-stealth] (blue) --  (green);
\draw[-stealth] (green) -- (yellow);
\draw[-stealth] (green) -- (red);
\draw[-stealth] (yellow) -- (red);


\end{tikzpicture}
\vspace{0.5cm}
\subcaption{This diagram gives a schematic representation of the boost transformations of the different torsion components. Arrows display relations between the different torsion components under boost transformations. If the boost transformation of a torsion component contains another torsion component, an arrow points from the former component towards the latter, e.g. the upmost arrow represents the boost transformation $\delta T^{(m)}_{\mu\nu} = \lambda_a{T_{\mu\nu}}^a$.}
\label{fig:particleconsistentboosttransform1a}
\end{subfigure}
\hfill
\begin{subfigure}[t]{0.45\textwidth}
\centering

\begin{tikzpicture}[scale=0.6,every node/.style={scale=0.6}]
\node at (0,0) {$5$};
\rectfour{1}{0}{red}{yellow}{green}{blue}
\node at (0,2) {$4$};
\rectfour{1}{2}{white}{yellow}{green}{blue}
\node at (0,4) {$3$};
\rectfour{1}{4}{white}{white}{green}{blue}
\node at (0,6) {$2$};
\rectfour{1}{6}{white}{white}{white}{blue}
\node at (0,8) {$1$};
\rectfour{1}{8}{white}{white}{white}{white}

\draw [-stealth] (2.5,7.5) -- (2.5,6.5);
\draw [-stealth] (2.5,5.5) -- (2.5,4.5);
\draw [-stealth] (2.5,3.5) -- (2.5,2.5);
\draw [-stealth] (2.5,1.5) -- (2.5,0.5);
\end{tikzpicture}
\vspace{0.5cm}
\subcaption{The colours in every rectangle of this diagram represent the parts of the corresponding torsion tensors that are unconstrained. In particular, if a colour is missing, it implies that this part of the torsion tensor is zero. In Figure \ref{fig:particleconsistentboosttransform1a}, we have indicated what colour corresponds to what part of the torsion tensors. Furthermore, arrows point from more special cases towards more general cases.}
\label{fig:particleconsistentboosttransform1b}
\end{subfigure}
\caption{Classification of constraints on the torsion tensors (b) that are consistent with the local structure group transformations (a).}
\label{fig:particleconsistentboosttransform}
\end{figure}

\noindent In the previous section, we saw that the specification of a generic torsionful affine connection that is compatible with the NC metric structure involves the introduction of time-like and spatial torsion tensors $T_{\mu\nu}$ and $T_{\mu\nu}{}^a$, as well as an extra mass torsion tensor $T^{(m)}_{\mu\nu}$. While we have thus far kept these tensors completely arbitrary, it is possible to consider special cases, in which some of their components are equal to zero. Since the components of $T_{\mu\nu}$, ${T_{\mu\nu}}^a$ and $T^{(m)}_{\mu\nu}$ transform non-trivially into each other under Galilean boosts, in a way that is summarized in Figure \ref{fig:particleconsistentboosttransform1a}, one cannot set their components equal to zero independently.

Let us illustrate this by outlining several scenarios in which components of the torsion tensors are set to zero consistently. These possible truncations are displayed in Figure \ref{fig:particleconsistentboosttransform1b}, where we have used the following notation to denote torsion tensor components:
\begin{align}
  T_{0a} = \tau^\mu e_a{}^\nu T_{\mu\nu} \,, \qquad \qquad T_{ab} = e_a{}^\mu e_b{}^\nu T_{\mu\nu} \,,
\end{align}
and similarly for components of $T_{\mu\nu}{}^a$ and $T^{(m)}_{\mu\nu}$. The cases displayed in Figure \ref{fig:particleconsistentboosttransform1b} can be retrieved from Figure \ref{fig:particleconsistentboosttransform1a} as follows: every possible scenario (a rectangle in Figure \ref{fig:particleconsistentboosttransform1b}) corresponds to setting the torsion components whose colour (indicated in Figure \ref{fig:particleconsistentboosttransform1a}) is absent, to zero. For example, case 4 of Figure \ref{fig:particleconsistentboosttransform1b} corresponds to setting $T_{ab}$ equal to zero. For consistency, it is then required that torsion components that are set to zero point towards torsion components that are also put equal to zero in Figure \ref{fig:particleconsistentboosttransform1a}. E.g., since in case 2 of Figure \ref{fig:particleconsistentboosttransform1b} $T_{\mu\nu}{}^a$ is set to zero, the components $T_{0a}$ and $T_{ab}$ also have to vanish.

\begin{figure}[t]
    \centering
    \includegraphics[width=.8\textwidth]{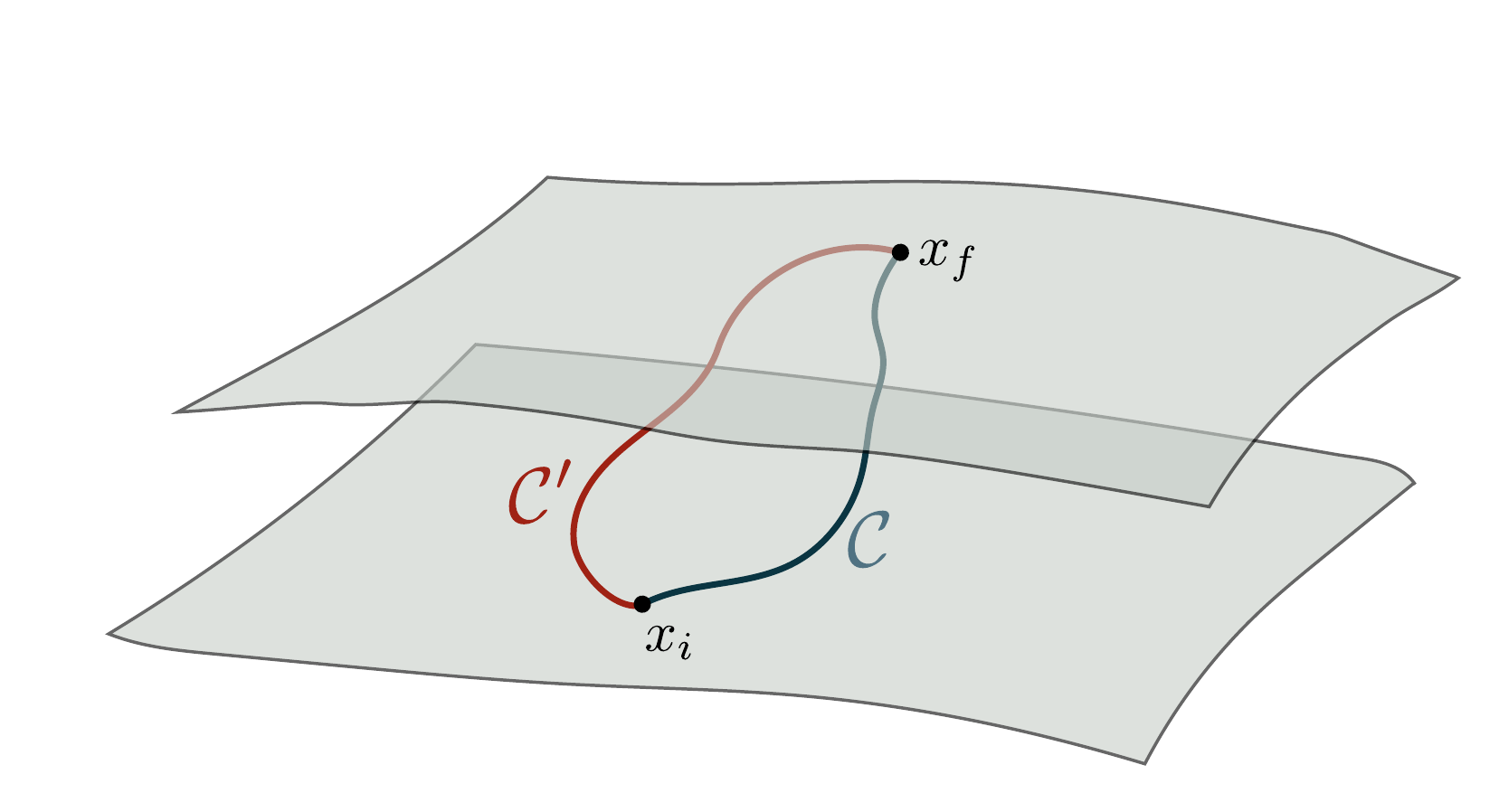}
    \caption{This figure depicts two events $x_i$, $x_f$ in space-time that are connected by two distinct future-directed time-like curves $\mathcal C$, $\mathcal C'$. With $\tau = \tau_\mu \rmd x^\mu$, Stokes' theorem implies that $\int_{\mathcal C}\tau - \int_{\mathcal C'}\tau = \int_{\partial \Sigma} \tau = \int_\Sigma \rmd \tau$, where $\Sigma$ is a surface enclosed by $\mathcal{C}$ and $\mathcal{C}'$. In case the time-like torsion $T_{\mu\nu}$ and thus $\rmd \tau$ are zero, one then finds that the time interval \eqref{eq:timeinterval} is path-independent: $\int_{\mathcal C}\tau = \int_{\mathcal C'}\tau$.}
    \label{fig:AbsoluteTime}
\end{figure}

A useful way to divide the different cases of Figure \ref{fig:particleconsistentboosttransform1b} is according to the following list.
\begin{itemize}
\item Cases 1 and 2 in Figure \ref{fig:particleconsistentboosttransform1b} correspond to the cases in which the affine connection has zero torsion ($T_{\mu\nu} = 0 = T_{\mu\nu}{}^a$). Case 1 is known as `torsionless NC geometry' in the literature. \footnote{The existence of case 2 is related to a difference between NC and Lorentzian geometry, namely the fact that in the former $\Gamma_{\mu\nu}^\rho$ is not uniquely specified by metric compatibility \eqref{eq:metricscomp} and $\Gamma^\rho_{[\mu\nu]} =0$, but only up to an ambiguity parametrized by a two-form $K_{\mu\nu}$. In case 1, this two-form is taken to be the curl of $m_\mu$. Case 2 indicates that $K_{\mu\nu}$ does not necessarily have to be exact and can instead be a generic two-form.}
\item Cases 1, 2 and 3 in Figure \ref{fig:particleconsistentboosttransform1b} have zero time-like/intrinsic torsion: $T_{\mu\nu} = 0$. Of these, case 3 has unconstrained (spatial) torsion of $\Gamma^\rho_{\mu\nu}$ (i.e., has unconstrained components of $T_{\mu\nu}{}^a$). The vanishing of $T_{\mu\nu}$ means that the time-like Vielbein $\tau_\mu$ is closed:
  \begin{align} \label{eq:zerointrinsictorsion}
    \partial_{[\mu} \tau_{\nu]} = 0 \,.
  \end{align}
  As illustrated in Figure \ref{fig:AbsoluteTime}, Stokes' theorem then implies that the time interval \eqref{eq:timeinterval} is independent of the curve that connects two particular events. Different physical observers that move along different curves between the same initial and final events thus measure the same time interval for their respective journeys. In other words, NC manifolds with vanishing intrinsic torsion admit a notion of absolute time. Locally, \eqref{eq:zerointrinsictorsion} implies that $\tau_\mu$ is exact, i.e. $\tau_\mu = \partial_\mu t$, and the function $t$ can be identified as an absolute time function.
\item Case 4 has $T_{ab} = 0$ but $T_{0a}$ unconstrained. These conditions are equivalent to stating that $\tau_\mu$ is hypersurface orthogonal:
  \begin{align} \label{eq:hypsurforth}
    \tau_{[\mu} \partial_\nu \tau_{\rho]} = 0 \,,
  \end{align}
  but not necessarily closed. This case is known as `twistless torsionful NC geometry' \cite{Christensen:2013lma,Christensen:2013rfa}. As can be seen from Figure \ref{fig:particleconsistentboosttransform1b}, consistency with Galilean boosts requires that both $T_{\mu\nu}{}^a$ and $T^{(m)}_{\mu\nu}$ cannot be set to zero in general. Unlike the previous cases, it is no longer possible to define an absolute time in twistless torsionful NC geometry, since the time interval \eqref{eq:timeinterval} between two particular events now depends on the path that connects them. There however still is a notion of absolute simultaneity. This follows from Frobenius' theorem, according to which a NC manifold on which the hypersurface orthogonality condition \eqref{eq:hypsurforth} holds, can be foliated in $(D-1)$-dimensional spatial hypersurfaces, i.e. hypersurfaces of simultaneous events. Locally, $\tau_\mu$ is only exact after multiplication with an integrating factor $\rme^{-\phi}$, i.e., one can write $\tau_\mu = \rme^{\phi} \partial_\mu t$. The $(D-1)$-dimensional leaves of the foliation are then given by the $t=\mathrm{constant}$ hypersurfaces. Twistless torsionful NC geometry then still has a notion of Newtonian causality in the sense that, given a spatial hypersurface $t=c_0$, one can distinguish its future, given by the collection of hypersurfaces $t=c_1$ with $c_1 > c_0$, from its past, given by the collection of hypersurfaces $t=c_2$ with $c_2 < c_0$.
  \item Case 5 leaves both $T_{0a}$ and $T_{ab}$ unconstrained. Consistency with boosts then requires that all torsion tensors are unconstrained. In this case, there is neither a notion of absolute time nor of absolute simultaneity and Newtonian causality.
\end{itemize}

Note that we have not given a complete classification of all possible scenarios in which the torsion components can be set to zero consistently. One could for instance split the torsion tensor $T^{(m)}_{\mu\nu}$ up into a part that (partially) projects on the longitudinal Vielbein $T_{0a}^{(m)}$ and a part that does not, $T_{ab}^{(m)}$, which would lead to a finer classification. We have also not considered cases, in which combinations of components of different torsion tensors are put equal to zero. 

Torsionless Newton-Cartan geometry is the geometry underlying Newton-Cartan gravity, the diffeomorphism covariant formulation of Newtonian gravity \cite{Cartan1,Cartan2}. Torsionful Newton-Cartan geometry has appeared in recent applications. Let us mention two examples. The first example deals with supergravity versions of Newton-Cartan gravity, that have thus far only been constructed in three space-time dimensions \cite{Andringa:2013mma,Bergshoeff:2015uaa,Bergshoeff:2015ija,Bergshoeff:2016lwr,Ozdemir:2019orp,deAzcarraga:2019mdn,Ozdemir:2019tby,Concha:2019mxx,Concha:2020tqx,Concha:2020eam,Concha:2021jos,Concha:2021llq}. These theories are based on torsionful NC geometry, where the torsion tensors $T_{\mu\nu}$, $T_{\mu\nu}{}^a$ and $T^{(m)}_{\mu\nu}$ are built out of fermionic gravitino fields. For example, the three-dimensional Newton-Cartan supergravity theory with 4 supercharges of \cite{Andringa:2013mma} contains two gravitino fields $\psi_{\mu+}$ and $\psi_{\mu-}$ that are both Majorana vector-spinors. Their transformation rules under local spatial rotations and Galilean boosts are given by
\begin{align} \label{eq:trafosgravitini}
  \delta \psi_{\mu+} = \frac14 \lambda^{ab} \gamma_{ab} \psi_{\mu+} \,, \qquad \qquad \delta \psi_{\mu-} = \frac14 \lambda^{ab} \gamma_{ab} \psi_{\mu-} - \frac12 \lambda^a \gamma_{a0} \psi_{\mu+} \,.
\end{align}
Here $\gamma_{ab} = \gamma_{[a} \gamma_{b]}$, $\gamma_{a0} = \gamma_a \gamma_0$ and $\{\gamma_0, \gamma_a | a=1,2\}$ constitute a set of three-dimensional gamma matrices (for a Clifford algebra with signature $(-++)$). The NC geometry used in \cite{Andringa:2013mma} then belongs to case 5 of Figure \ref{fig:particleconsistentboosttransform1b}, with torsion tensors $T_{\mu\nu}$, $T_{\mu\nu}{}^a$ and $T^{(m)}_{\mu\nu}$ constructed out of $\psi_{\mu\pm}$ as follows:
\begin{align}
  T_{\mu\nu} = \frac12 \bar{\psi}_{[\mu+} \gamma^0 \psi_{\nu]+} \,, \qquad \qquad T_{\mu\nu}{}^a = \bar{\psi}_{[\mu+} \gamma^a \psi_{\nu]-} \,, \qquad \qquad T^{(m)}_{\mu\nu} = \bar{\psi}_{[\mu-} \gamma^0 \psi_{\nu]-} \,.
\end{align}
Using the transformation rules \eqref{eq:trafosgravitini} of $\psi_{\mu\pm}$, one then finds that these torsion tensors satisfy the transformation rules \eqref{eq:trafoTTa} and \eqref{eq:trafoTM} that are required for invariance of the affine connection $\Gamma^\rho_{\mu\nu}$ \eqref{eq:Gammaexpr} under local rotations and boosts. Note however that $\Gamma^\rho_{\mu\nu}$ is not invariant under supersymmetry.

Our second example concerns NC geometry as it occurs in attempts to extend the AdS/CFT correspondence to describe non-relativistic conformal field theories (CFTs) \cite{Son:2008ye,Balasubramanian:2008dm,Kachru:2008yh}. In these proposals, non-relativistic CFTs live on the boundary of so-called Schr\"odinger or Lifshitz space-times that are vacuum solutions of matter coupled relativistic bulk gravity theories, and whose isometries form a non-relativistic conformal symmetry group. CFT quantities are then holographically encoded in bulk gravitational ones. While Schr\"odinger or Lifshitz space-times are relativistic in the bulk, their boundaries have a non-relativistic causal structure and are thus naturally described by NC geometry. It has in particular been shown that in holography around Lifshitz space-times, the relevant boundary geometry is that of torsionful NC geometry \cite{Christensen:2013lma,Christensen:2013rfa} in which the intrinsic torsion is non-vanishing (as in cases 4 and 5 in Figure \ref{fig:particleconsistentboosttransform1b}). The torsion tensors $T_{\mu\nu}$, $T_{\mu\nu}{}^a$ and $T^{(m)}_{\mu\nu}$ that occur are expressed in terms of the frame fields $\tau_\mu$, $e_\mu{}^a$ and $m_\mu$ and possible choices are given by\cite{Hartong:2015zia}:
\begin{align}
  \label{eq:deptorsion1}
  T_{\mu\nu} = 2 \partial_{[\mu} \tau_{\nu]} \,, \qquad \qquad T_{\mu\nu}{}^a = 2 e^{a\rho} m_\rho \partial_{[\mu} \tau_{\nu]} \,, \qquad \qquad T^{(m)}_{\mu\nu} = e^{a\rho} e_a{}^\sigma m_\rho m_\sigma \partial_{[\mu} \tau_{\nu]} \,,
\end{align}
and
\begin{align}
  \label{eq:deptorsion2}
  T_{\mu\nu} = 2 \partial_{[\mu} \tau_{\nu]} \,, \qquad \qquad  T_{\mu\nu}{}^a = 2 e^{a\rho} m_\rho \partial_{[\mu} \tau_{\nu]} \,, \qquad \qquad T^{(m)}_{\mu\nu} = - 2 \tau^\rho  m_\rho \partial_{[\mu} \tau_{\nu]} \,.
\end{align}
Using the rules \eqref{eq:localtrafosindep} and \eqref{eq:localtrafosinv}, one sees that these tensors indeed transform under the structure group as in \eqref{eq:trafoTTa} and \eqref{eq:trafoTM}. Both torsion tensor choices of \eqref{eq:deptorsion1}, \eqref{eq:deptorsion2} are, however, not invariant under the central charge transformation \eqref{eq:centralcharge}. Consequently, the affine connections constructed using them are invariant under local rotations and boosts, but not under the central charge transformation. As a result, central charge invariance is usually only realized in a non-manifest manner in holographic descriptions of non-relativistic CFTs.

\section{Torsionful String Newton-Cartan geometry in the Cartan formulation} \label{sec:TNC2}

\noindent In the previous section, we described NC geometry, which forms the natural differential geometric arena for non-relativistic particle mechanics. The framework of NC geometry can be generalized to manifolds, in which one can describe the movement of extended objects, such as strings and branes, in a degenerate limit that is akin to a non-relativistic one. Here we will focus on so-called non-relativistic strings  \cite{Gomis:2000bd,Danielsson:2000gi,Danielsson:2000mu} (see also \cite{Oling:2022fft} for a recent review). These are obtained from relativistic strings by sending the speed of light in the directions transversal to the strings to infinity, while leaving the relativistic character of the worldsheet untouched. Upon quantization, one then finds that this limit only retains vibrational modes with non-relativistic dispersion relations in the string spectrum. The target space-times that non-relativistic strings move in are referred to as String Newton-Cartan (SNC) manifolds and their geometry is likewise called SNC geometry.

Similar to NC geometry, $D$-dimensional SNC manifolds have a degenerate metric structure that reduces the local structure group to
\begin{align}
  \label{eq:structgroupSNC}
  \left( \mathrm{SO}(1,1) \times \mathrm{SO}(D-2)\right) \rtimes \mathbb{R}^{2(D-2)} \,.
\end{align}
The Minkowskian worldsheet of a non-relativistic string at rest divides up the tangent space directions of a SNC manifold in two `longitudinal' directions and $D-2$ `transversal' ones. The SO$(1,1)$ and SO$(D-2)$ factors of the structure group then correspond to Lorentz transformations of the two longitudinal directions and rotations of the transversal directions, respectively. The $\mathbb{R}^{2(D-2)}$ factor represents boost transformations that can transform transversal directions into longitudinal ones, but not vice versa. We will refer to these as `String Galilean boosts'. In the Lie algebra of \eqref{eq:structgroupSNC} the generators of $\mathbb{R}^{2(D-2)}$ then transform in the $(\mathbf{2},\mathbf{D-2})$ representation under the adjoint action of the Lie algebras of SO$(1,1)$ and SO$(D-2)$.

For the torsionless case, a Cartan formulation of SNC geometry was discussed from the viewpoint of space-time symmetry algebra gaugings and a particular limit of the Cartan formulation of Lorentzian geometry in \cite{Andringa:2012uz,Bergshoeff:2019pij}.\footnote{For earlier work on SNC geometry, see \cite{Gomis:2005pg,Brugues:2004an,Brugues:2006yd}}. Recently, the relevance of including non-trivial torsion in SNC geometry has been pointed out in \cite{Gallegos:2020egk,Bergshoeff:2021bmc,Bidussi:2021ujm,Bergshoeff:2021tfn}. In this section, we will present the metric and affine connection structure of torsionful SNC geometry, in the same spirit as our presentation of torsionful NC geometry of the previous section. We will first discuss the frame fields and resulting metric structure of the Cartan formulation of SNC geometry in section \ref{ssec:framemetricTSNC}. Next, in section \ref{ssec:connectionTSNC}, we will introduce a metric compatible affine connection by introducing suitable structure group spin connections and Vielbein postulates. As in the previous section, we will at first leave the torsion arbitrary and independent. We will see that, unlike what happens for NC geometry, the affine connection of SNC geometry can (for our choice of frame fields) no longer be fully expressed in terms of frame fields and independent torsion tensors. As in the NC case, it is possible to consider various special cases that are obtained by truncating torsion tensor components consistently. This will be treated in section \ref{ssec:specialTSNC}, with particular emphasis on cases that have appeared in the recent literature.

\subsection{Frame fields and metric structure} \label{ssec:framemetricTSNC}

In analogy to NC geometry, the Cartan formulation of SNC geometry includes three different types of frame fields: a `longitudinal Vielbein' $\tau_\mu{}^A$ ($A=0,1$), a `transversal Vielbein' $e_\mu{}^{a}$ ($a = 2, \cdots, D-1$)\footnote{Many articles on SNC-type geometries use primed capital letters $A',B',C',\cdots$ for the transversal directions instead of the lowercase $a,b,c,\cdots$ used here.} and a two-form field $b_{\mu\nu}$. The flat longitudinal index $A$ can be freely raised and lowered with a two-dimensional Minkowski metric $\eta_{AB} = \mathrm{diag}(-1,1)$, whereas for the flat transversal index $a$ this is done using a $(D-2)$-dimensional Euclidean metric $\delta_{ab}$. The frame fields transform under the structure group \eqref{eq:structgroupSNC} in a reducible, indecomposable manner according to the following local transformation rules:
\begin{align} \label{eq:localtrafosframeSNC}
  \delta \tau_\mu{}^A &= \lambda_M \epsilon^A{}_B \tau_\mu{}^B \,, \qquad \qquad \qquad \delta e_\mu{}^{a} = \lambda^{a}{}_{b} e_\mu{}^{b} - \lambda_A{}^{a} \tau_\mu{}^A \,, \nonumber \\
  \delta b_{\mu\nu} &= -2 \epsilon_{AB} \lambda^A{}_{a} \tau_{[\mu}{}^B e_{\nu]}{}^{a} \,.
\end{align}
Here, $\lambda_M$ corresponds to the parameter of longitudinal SO$(1,1)$ Lorentz transformations, $\lambda^{ab} = - \lambda^{ba}$ to that of transversal SO$(D-2)$ rotations, while the $\lambda^{Aa}$ are the $2(D-2)$ String Galilean boost parameters. Note that the String Galilean boosts act in a non-linear fashion on the two-form field $b_{\mu\nu}$. Similar to NC geometry, one introduces an `inverse longitudinal Vielbein' $\tau_A{}^\mu$ and an `inverse transversal Vielbein' $e_{a}{}^\mu$ via the following equations:
\begin{alignat}{3}
  \label{eq:invVielbeineSNC}
  & \tau_A{}^\mu \tau_\mu{}^B = \delta_A^B \,, \qquad \qquad \qquad & & \tau_A{}^\mu e_\mu{}^{a} = 0 \,, \qquad \qquad \qquad \qquad & & e_{a}{}^\mu \tau_\mu{}^A = 0 \,, \nonumber \\
  & e_\mu{}^{a} e_{b}{}^\mu = \delta_{b}^{a} \,, \qquad \qquad \qquad & & \tau_\mu{}^A \tau_A{}^\nu + e_\mu{}^{a} e_{a}{}^\nu = \delta_\mu^\nu \,.
\end{alignat}
These relations express that the matrices $\begin{pmatrix} \tau_\mu{}^A & e_\mu{}^{a}\end{pmatrix}$ and $\begin{pmatrix} \tau_A{}^\mu \\ e_{a}{}^\mu\end{pmatrix}$ are each other's inverse. The transformation rules of $\tau_A{}^\mu$ and $e_{a}{}^\mu$ under the action of the local structure group are then given by:
\begin{align}
  \label{eq:localtrafosinvSNC}
  \delta \tau_A{}^\mu &= \lambda_M \epsilon_A{}^B \tau_B{}^\mu + \lambda_{A}{}^{a} e_{a}{}^\mu \,, \qquad \qquad \qquad \delta e_{a}{}^\mu = \lambda_{a}{}^{b} e_{b}{}^\mu \,.
\end{align}

The longitudinal and inverse transversal Vielbeine can be `squared' to obtain two degenerate symmetric (covariant and contravariant) two-tensors that are invariant under local SO$(1,1)$, SO$(D-2)$ and String Galilean boost transformations:
\begin{align} \label{eq:TSNCmetricstruct}
  \tau_{\mu\nu} \equiv \tau_\mu{}^A \tau_\nu{}^B \eta_{AB} \,, \qquad \qquad \qquad  h^{\mu\nu} \equiv e_{a}{}^{\mu} e_{b}{}^\nu \delta^{ab} \,.
\end{align}
These two tensors constitute a degenerate metric structure on a SNC manifold. The covariant metric $\tau_{\mu\nu}$ is referred to as the `longitudinal metric'. From \eqref{eq:invVielbeineSNC} one sees that its kernel is spanned by the $D-2$ vectors $e_{a}{}^\mu$ and it thus has rank 2. The contravariant metric $h^{\mu\nu}$ is called the `transversal metric' and has rank $D-2$, since its kernel is spanned by the two one-forms $\tau_\mu{}^A$.

\begin{figure}[t]
  \centering
  \includegraphics[width=.8\textwidth]{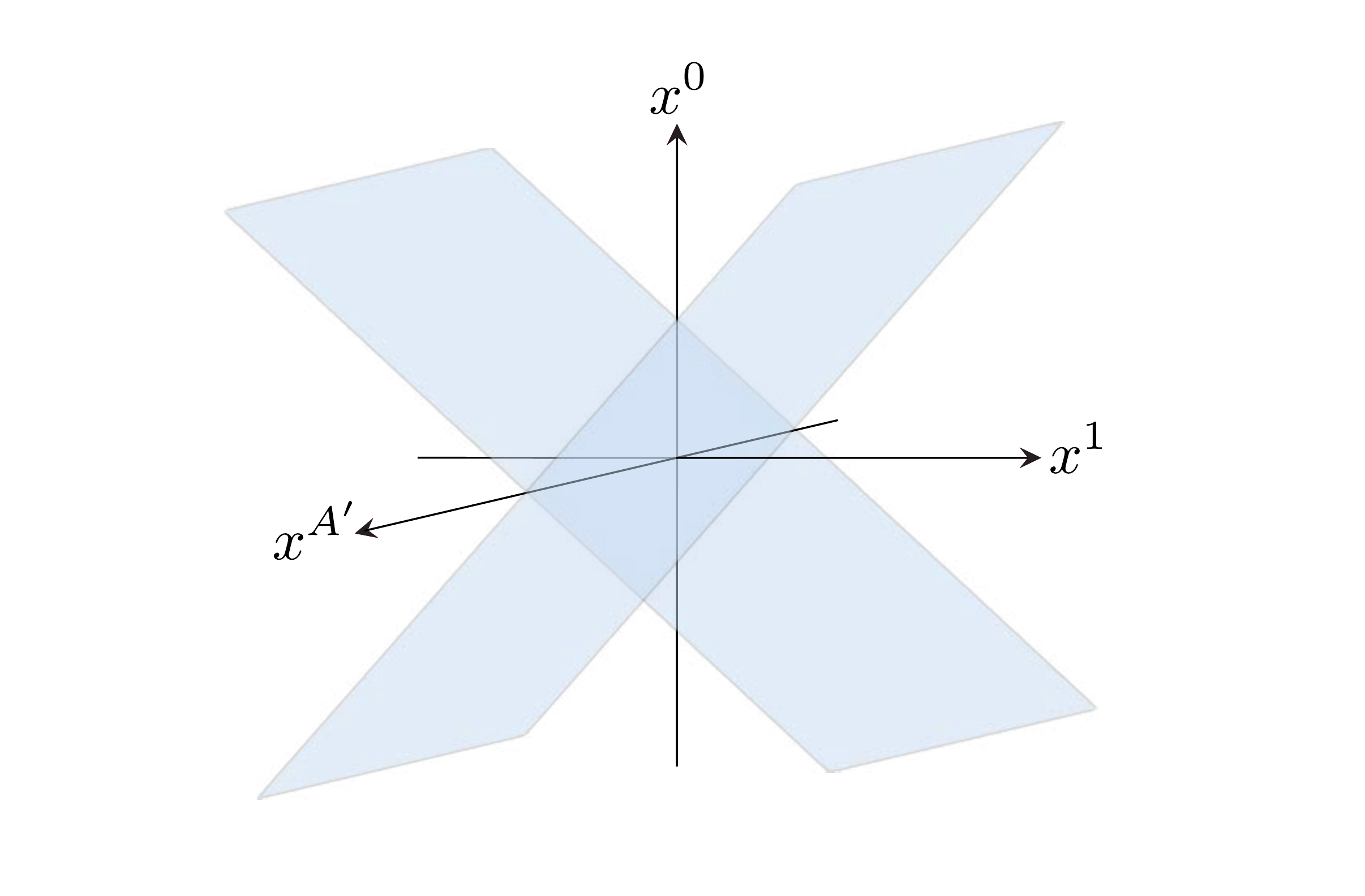}
  \caption{Local causal structure of an SNC manifold. The Lorentzian lightcone degenerates into a ``lightwedge'', defined by the two hyperplanes $x^0\pm x^1=0$.}
  \label{fig:Lightwedge}
\end{figure}

Similar to NC geometry, the local causal structure of a SNC manifold can be obtained as a degenerate limit of that of a Lorentzian manifold. In this case, this limit consists of sending the velocity of light in the transverse directions in a local inertial reference frame to infinity. In local Minkowski coordinates $x^{\hat{A}}=\{x^A, x^{a}\}$ (with $\hat{A} = 0,\cdots, D-1)$, this is achieved by rescaling the longitudinal coordinates $x^A$ with a (dimensionless) parameter $\omega$ and taking the limit $\omega \rightarrow \infty$. The local lightcone $\omega^2 x^A x_A = -x^{a} x_{a}$ then flattens out along the transversal directions and degenerates into the two hyperplanes $x^0 = x^1$ and $x^0 = - x^1$; see figure \ref{fig:Lightwedge}. A vector can then be distinguished according to whether it lies in the $(D-2)$-dimensional intersection of these two hyperplanes or not. In the former case, we will call the vector `transversal', while in the latter case we will call it a `worldsheet vector'. Worldsheet vectors can be further classified as time-like, space-like or null vectors, according to whether their projections onto the $(x^0, x^1)$-plane is time-like, space-like or null with respect to the two-dimensional Minkowski metric $\eta_{AB}$. Put covariantly, a vector $X^\mu$ is transversal whenever $\tau_\mu{}^A X^\mu = 0$ for $A=0,1$ and a worldsheet vector whenever $\tau_\mu{}^A X^\mu$ are not both zero. Distinguishing worldsheet vectors into time-like, space-like or null ones is done using the longitudinal metric $\tau_{\mu\nu}$. In particular, a worldsheet vector $X^\mu$ is time-like whenever $\tau_{\mu\nu} X^\mu X^\nu = \tau_\mu{}^A X^\mu \tau_{\nu A} X^\nu < 0$, space-like whenever $\tau_{\mu\nu} X^\mu X^\nu  > 0$ and null whenever $\tau_{\mu\nu} X^\mu X^\nu = 0$.

Given a SNC manifold $\mathcal{M}$, the longitudinal metric $\tau_{\mu\nu}$ can be used to calculate a proper time
\begin{align}
    \int_0^1 \rmd \tau \sqrt{- \tau_{\mu \nu} \dot{x}^\mu \dot{x}^\nu} \,,
\end{align}
along a curve segment $\gamma : \tau \in [0,1]  \rightarrow x^\mu(\tau) \in \mathcal{M}$, for which $\dot{x}^\mu(\tau) \equiv \rmd x^\mu(\tau)/\rmd \tau$ is a time-like worldsheet vector for all $\tau \in (0,1)$. Similarly, if $\tilde{\gamma} : \sigma \in [0,1] \rightarrow x^\mu(\sigma) \in \mathcal{M}$ is a curve segment, for which $x^{\prime \mu}(\sigma) \equiv \rmd x^\mu(\sigma)/\rmd \sigma$ is a space-like worldsheet vector for all $s \in (0,1)$, one can define its proper length as
\begin{align}
    \int_0^1 \rmd \sigma \sqrt{\tau_{\mu \nu} x^{\prime \mu} x^{\prime \nu}} \,.
\end{align}
Furthermore, $\tau_{\mu\nu}$ can also be used to give a notion of proper area of worldsheets, whose tangent vectors are worldsheet vectors. In particular, the proper area of a worldsheet segment $\Sigma_\varphi$ that is specified via an embedding map $\varphi: (\tau,\sigma) \in [0,1] \times [0, 2 \pi] \rightarrow x^\mu(\tau,\sigma) \in \mathcal{M}$, such that $\partial_\tau x^\mu(\tau,\sigma)$ and $\partial_\sigma x^\mu(\tau,\sigma)$ are time-like, resp. space-like worldsheet vectors, can be defined as:
\begin{align}
    \label{eq:properarea1}
    \int_0^1 \rmd \tau \int_0^{2\pi} \rmd \sigma \, \sqrt{-\mathrm{det}(\tau_{\alpha\beta})} \,, \qquad \qquad \text{with} \ \ \ \tau_{\alpha\beta} = \tau_{\mu\nu} \partial_\alpha x^\mu \partial_\beta x^\nu \,,
\end{align}
where the indices $\alpha$, $\beta$ can stand for $\tau$ or $\sigma$. Assuming that $\epsilon_{AB} \tau_\mu{}^A \partial_\tau x^\mu \tau_\nu{}^B \partial_\sigma x^\nu > 0$ for all possible values of $\tau$ and $\sigma$, where $\epsilon_{AB}$ is the two-dimensional Levi-Civita epsilon symbol, normalized as $\epsilon_{01} = 1$. Then the proper area can alternatively be written as the integral of the pullback of a two-form $\ell_{\mu\nu}$ over $\Sigma_\varphi$:
\begin{align}\label{eq:properarea2}
    2\,\int_0^1 \rmd \tau \int_0^{2\pi} \rmd \sigma \, \ell_{\mu\nu} \partial_\tau x^\mu \partial_\sigma x^\nu \,,\qquad \qquad \text{with} \ \ \ \ell_{\mu\nu} = \frac12\,\epsilon_{AB}\tau_\mu{}^A\tau_\nu{}^B\,.
\end{align}
This is analogous to how the time interval \eqref{eq:timeinterval} in NC geometry can be given by integrating the pullback of the one-form $\tau_\mu$ along a worldline. Note that this notion of proper worldsheet area does not exist in NC geometry, since there the only metric that can act on time-like vectors is of rank 1.

The rank $D-2$ co-metric $h^{\mu\nu}$ can be used to measure transversal distances to worldsheets. To do this, one proceeds similarly as in NC geometry and one views $h^{\mu\nu}$ as a well-defined and invertible map between the space of equivalence classes $[\alpha_\mu] = \{\alpha_\mu + f_A \tau_\mu{}^A | f_A \in C^\infty(\mathcal{M})\}$ of one-forms that differ by linear combinations of $\tau_\mu{}^A$ and the space of transversal vectors, where $h^{\mu\nu}$ maps $[\alpha_\nu]$ to $h^{\mu\nu}[\alpha_\nu] \equiv h^{\mu\nu} \alpha_\nu$. In analogy to the NC geometry case, one can argue that the inverse of this map is given by:
\begin{align}
  h_{\mu\nu} = e_\mu{}^{a} e_\nu{}^{b} \delta_{ab} \,,
\end{align}
where one regards $h_{\mu\nu}$ as a map that assigns the equivalence class $[h_{\mu\nu} X^\nu]$ to each transversal vector $X^\mu$. 
Note that $h_{\mu\nu}$ cannot be viewed as a covariant metric on the full space of vectors, since it is not invariant under local boosts: $\delta h_{\mu\nu} = -2 \lambda_A{}^{a} \tau_{(\mu}{}^A e_{\nu) a}$. It does however form a covariant metric (with Euclidean signature) on the space of transversal vectors, since $X^\mu Y^\nu h_{\mu\nu}$ is boost invariant when $X^ \mu$ and $Y^\mu$ are transversal\footnote{As in the NC case, a slightly stronger statement is that the equivalence class $[h_{\mu\nu} X^\nu]$ is boost invariant, when $X^\mu$ is transversal.}. One can thus use it to define a transversal distance notion along any curve $s \in [0,1] \rightarrow x^\mu(s)$, whose tangent vectors $x^{\prime \mu}(s) \equiv \rmd x^\mu(s)/\rmd s$ are transversal for all $s \in (0,1)$ as follows:
\begin{align}
  \label{eq:length2}
  \int_0^1 \rmd s \, \sqrt{x^{\prime \mu} x^{\prime \nu} h_{\mu\nu}} \,.
\end{align}
The frame fields $\tau_\mu{}^A$ and $e_\mu{}^{a}$ are a natural generalization of the time-like and spatial Vielbeine $\tau_\mu$ and $e_\mu{}^a$ of NC geometry. The frame field $b_{\mu\nu}$ plays a very similar role in SNC geometry as the mass form $m_\mu$ does in NC geometry. It is not needed to specify the metric structure on the NC geometry, but it becomes part of the definition of a metric compatible affine connection in terms of frame fields and torsion tensors, as we will review in the next section.

\subsection{Torsionful, metric compatible connection} \label{ssec:connectionTSNC}

\noindent To define a metric compatible affine connection in SNC geometry, we proceed analogously as in NC geometry and first introduce a structure group connection $\Omega_\mu$ that takes values in the Lie algebra of \eqref{eq:structgroupSNC}
\begin{equation}
  \Omega_\mu = \omega_\mu J + \frac12 \omega_\mu{}^{ab} J_{ab} + \omega_\mu{}^{A a} G_{A a} \,,
\end{equation}
where $J$, $J_{ab} = -J_{ba}$ and $G_{A a}$ are generators of the Lie algebras of SO$(1,1)$, SO$(D-2)$ and $\mathbb{R}^{2(D-2)}$. We will refer to $\omega_\mu$, $\omega_\mu{}^{ab} = - \omega_\mu{}^{ba}$ and $\omega_\mu{}^{A a}$ as spin connections for longitudinal Lorentz transformations, transversal rotations and String Galilean boosts, respectively. They transform as follows under infinitesimal SO$(1,1)$, SO$(D-2)$ and String Galilean boosts:
\begin{align}
  \label{eq:localtrafosdepSNC}
    \delta \omega_\mu &= \partial_\mu \lambda_M \,, \qquad \qquad
    \delta \omega_\mu{}^{ab} = \partial_\mu \lambda^{ab} + 2 \lambda^{[a|c|} \omega_{\mu c}{}^{b]} \,, \nonumber \\
  \delta \omega_\mu{}^{A a} &= \partial_\mu \lambda^{A a} + \lambda_M \epsilon^A{}_B \omega_\mu{}^{B a}+ \lambda^{a}{}_{b} \omega_\mu{}^{A b} - \epsilon^A{}_B \lambda^{B a} \omega_\mu + \lambda^{A b} \omega_{\mu b}{}^{a} \,.
\end{align}
An affine connection $\Gamma_{\mu\nu}^\rho$ can be introduced by imposing the following `Vielbein postulates':
\begin{align}
  \label{eq:VielbpostSNC}
  & \partial_\mu \tau_\nu{}^A - \epsilon^A{}_B \omega_\mu \tau_\nu{}^B -  \Gamma_{\mu\nu}^\rho \tau_\rho{}^A = 0 \,, \nonumber \\
  &  \partial_\mu e_\nu{}^{a} - \omega_\mu{}^{ab} e_{\nu b} + \omega_\mu{}^{A a} \tau_{\nu A} - \Gamma_{\mu\nu}^\rho e_{\rho}{}^{a} = 0 \,.
\end{align}
These postulates imply that $\Gamma_{\mu\nu}^\rho$ is compatible with the SNC metric structure \eqref{eq:TSNCmetricstruct}:
\begin{align}
  \label{eq:metricscompSNC}
  \nabla_\mu \tau_{\nu\rho } & \equiv \partial_\mu \tau_{\nu\rho} - \Gamma_{\mu\nu}^\sigma \tau_{\sigma \rho} - \Gamma_{\mu\rho}^\sigma \tau_{\nu \sigma} = 0 \,, \qquad \qquad \nabla_\mu h^{\nu\rho}  \equiv \partial_\mu h^{\nu\rho} + \Gamma_{\mu\sigma}^\nu h^{\sigma \rho} + \Gamma_{\mu\sigma}^\rho h^{\nu\sigma}= 0 \,.
\end{align}
The connection $\Gamma^\rho_{\mu\nu}$ is assumed to transform appropriately under diffeomorphisms and to be invariant under the local structure group, so that the Vielbein postulates \eqref{eq:VielbpostSNC} are covariant with respect to all these transformations. In particular, under String Galilean boosts the first postulate does not transform, while the second one is boosted to the first one. Using \eqref{eq:VielbpostSNC}, one can express $\Gamma_{\mu\nu}^\rho$ in terms of the Vielbeine $\tau_\mu{}^A$, $e_\mu{}^{a}$, their inverses and the spin connections $\omega_\mu$, $\omega_\mu{}^{ab}$, $\omega_\mu{}^{A a}$:
\begin{align} \label{eq:GammaVielbpostSNC}
  \Gamma^\rho_{\mu\nu}  = \tau_A{}^\rho \partial_\mu \tau_\nu{}^A + e_{a}{}^\rho \partial_\mu e_\nu{}^{a} - \epsilon^A{}_B \omega_\mu \tau_\nu{}^B \tau_A{}^\rho - \omega_\mu{}^{a}{}_{b} e_{\nu}{}^{b} e_{a}{}^\rho + \omega_\mu{}^{A a} \tau_{\nu A} e_{a}{}^\rho \,.
\end{align}

As in the previous section, we will view the torsion $2 \Gamma_{[\mu\nu]}^\rho$ of the affine connection as an independent and a priori arbitrary geometric ingredient. We will split it into `longitudinal torsion' components $T_{\mu\nu}{}^A$ along $\tau_A{}^\rho$ and `transversal torsion' components $T_{\mu\nu}{}^{a}$ along $e_{a}{}^\rho$:
\begin{align} \label{eq:torsiondecompSNC}
  & 2 \Gamma^\rho_{[\mu\nu]} = \tau_A{}^\rho T_{\mu\nu}{}^A + e_{a}{}^\rho T_{\mu\nu}{}^{a} \qquad \text{i.e.} \qquad
    T_{\mu\nu}{}^A  \equiv 2 \Gamma^\rho_{[\mu\nu]} \tau_\rho{}^A  \quad \text{and} \quad T_{\mu\nu}{}^{a} = 2 \Gamma^\rho_{[\mu\nu]} e_\rho{}^{a} \,.
\end{align}
Under local SO$(1,1)$, SO$(D-2)$ and String Galilean boosts, $T_{\mu\nu}{}^A$ and $T_{\mu\nu}{}^{a}$ then transform as follows
\begin{align}
  \label{eq:trafoTATAp}
  \delta T_{\mu\nu}{}^A = \lambda_M \epsilon^A{}_B T_{\mu\nu}{}^B \,, \qquad \qquad \qquad \delta T_{\mu\nu}{}^{a} = \lambda^{a}{}_{b} T_{\mu\nu}^{b} - \lambda_A{}^{a} T_{\mu\nu}{}^A \,.
\end{align}
By antisymmetrizing the Vielbein postulates \eqref{eq:VielbpostSNC}, one obtains the following equations that are covariant with respect to local structure group transformations:
\begin{align}
  T_{\mu\nu}{}^A &= 2 \partial_{[\mu} \tau_{\nu]}{}^A - 2 \epsilon^A{}_B \omega_{[\mu} \tau_{\nu]}{}^B \,, \label{eq:torsionVielbSNC1} \\  T_{\mu\nu}{}^{a} &= 2 \partial_{[\mu} e_{\nu]}{}^{a} - 2 \omega_{[\mu}{}^{ab} e_{\nu] b} + 2 \omega_{[\mu}{}^{A a} \tau_{\nu] A} \label{eq:torsionVielbSNC2} \,.
\end{align}
The first of these represents a set of $D(D-1)$ equations. Of these, $D$ equations contain the $D$ components of $\omega_\mu$ algebraically, while the remaining $D(D-2)$ ones do not contain components of $\omega_\mu$. One can thus use $D$ of the equations \eqref{eq:torsionVielbSNC1} to express $\omega_\mu$ in terms of frame fields and components of the longitudinal torsion $T_{\mu\nu}{}^A$. Doing this leads to the following expression for $\omega_\mu$:
\begin{align}
  \label{eq:exprso11conn}
  \omega_\mu &= \epsilon^{AB} \tau_A{}^\nu \partial_{[\mu} \tau_{\nu] B} - \frac12 \epsilon^{BC} \tau_\mu{}^A \tau_B{}^\nu \tau_C{}^\rho \partial_{[\nu} \tau_{\rho]A} + \frac12 \epsilon^{BC} \tau_{\mu A} \tau_B{}^\nu \tau_C{}^\rho T_{\nu\rho}{}^A \nonumber \\ & \qquad + \frac12 \epsilon^A{}_B e_\mu{}^{a} \tau_A{}^\nu e_{a}{}^\rho T_{\nu \rho}{}^{B} \,.
\end{align}
The remaining $D(D-2)$ equations, contained in \eqref{eq:torsionVielbSNC1}, are given by:
\begin{align}
  \label{eq:intrinsictorsionSNC}
  \tau_{(A|}{}^\mu e_{a}{}^\nu T_{\mu\nu |B)} = 2 \tau_{(A|}{}^\mu e_{a}{}^\nu \partial_{[\mu} \tau_{\nu]|B)} \,, \qquad \qquad e_{a}{}^\mu e_{b}{}^\nu T_{\mu\nu}{}^A = 2 e_{a}{}^\mu e_{b}{}^\nu \partial_{[\mu} \tau_{\nu]}{}^A \,.
\end{align}
We thus see that only the components $\tau_B{}^\nu \tau_C{}^\rho T_{\nu\rho}{}^A$ and $\tau_{[A|}{}^\nu e_{a}{}^\rho T_{\nu \rho|B]}$ of the longitudinal torsion tensor $T_{\mu\nu}{}^A$ can be absorbed in the expression for a spin connection. The remaining components $\tau_{(A|}{}^\mu e_{a}{}^\nu T_{\mu\nu |B)}$ and $e_{a}{}^\mu e_{b}{}^\nu T_{\mu\nu}{}^A$ remain as intrinsic torsion and are involved in geometric constraints on the curl of $\tau_\mu{}^A$.

Equation \eqref{eq:torsionVielbSNC2} can be used to express some components of $\omega_\mu{}^{ab}$ and $\omega_\mu{}^{A a}$ in terms of frame fields and the transversal torsion tensor $T_{\mu\nu}{}^{a}$. Note that this cannot be done for all components of these spin connections, since \eqref{eq:torsionVielbSNC2} constitutes a set of $D(D-1)(D-2)/2$ equations, while there are $D(D+1)(D-2)/2$ components in $\omega_\mu{}^{ab}$ and $\omega_\mu{}^{A a}$. One can however use \eqref{eq:torsionVielbSNC2} to express the following $D(D-1)(D-2)/2$ spin connection components
\begin{align}
\label{eq:intrinsictorsionstring1}
  \tau_{[A|}{}^\mu \omega_{\mu|B]}{}^{a} \,, \qquad \tau_A{}^\mu \omega_\mu{}^{ab} \,, \qquad e_{(a|}{}^\mu \omega_{\mu A|b)} \,, \qquad e_{c}{}^\mu \omega_\mu{}^{ab}
\end{align}
in terms of frame fields, $T_{\mu\nu}{}^{a}$ and (some of) the remaining $D(D-2)$ components of $\omega_\mu{}^{ab}$ and $\omega_\mu{}^{A a}$. In order to also solve some of these remaining spin connection components in terms of frame fields and torsion, one can apply a similar strategy as in the NC geometry case. We thus introduce an extra independent torsion tensor $T^{(b)}_{\mu\nu\rho}$ and set it equal to the covariantized (with respect to String Galilean boosts) field strength of the two-form field $b_{\mu\nu}$:
\begin{align}
  \label{eq:Tisdb}
  T^{(b)}_{\mu\nu\rho} = 3 \partial_{[\mu} b_{\nu\rho]} + 6 \epsilon_{AB} \omega_{[\mu}{}^{A b} \tau_\nu{}^B e_{\rho] b} \,.
\end{align}
Note that the right-hand side of this equation is invariant under the following one-form gauge symmetry, with parameter $\sigma_\mu$:
\begin{align} \label{eq:oneformsymm}
\delta b_{\mu\nu} = 2 \partial_{[\mu} \sigma_{\nu]} \,.
\end{align}
We then find $T^{(b)}_{\mu\nu\rho}$ to be invariant under SO$(1,1)$ and SO$(D-2)$ transformations and to transform under String Galilean boosts as follows:
\begin{align}
  \label{eq:boostT}
  \delta T^{(b)}_{\mu\nu\rho} = -3 \epsilon_{AB} \lambda^A{}_{a} T_{[\mu\nu}{}^B e_{\rho]}{}^{a} + 3 \epsilon_{AB} \lambda^A{}_{a} T_{[\mu\nu}{}^{a} \tau_{\rho]}{}^B \,.
\end{align}
With this choice, the equations \eqref{eq:torsionVielbSNC1}, \eqref{eq:torsionVielbSNC2} and \eqref{eq:Tisdb} form an invariant set under \eqref{eq:localtrafosframeSNC}, \eqref{eq:localtrafosdepSNC}, \eqref{eq:trafoTATAp} and \eqref{eq:boostT}. As in the NC geometry case, this ensures that our final expressions for the spin connections in terms of frame fields and torsion tensors still transform as in \eqref{eq:localtrafosdepSNC}. Of the $D(D-1)(D-2)/6$ equations \eqref{eq:Tisdb}, $(D-2)^2$ equations can be used to express the following spin connection components
\begin{align}
  \tau_A{}^\mu \omega_\mu{}^{A a} \,, \qquad \qquad e_{[a|}{}^\mu \omega_{\mu A|b]} \,,
\end{align}
in terms of frame fields and $T^{(b)}_{\mu\nu\rho}$. The remaining $(D-2)(D-3)(D-4)/6$ equations take the form
\begin{align}
  \label{eq:intrinsicdb}
  e_{a}{}^\mu e_{b}{}^\nu e_{c}{}^\rho T^{(b)}_{\mu\nu\rho} = 3 e_{a}{}^\mu e_{b}{}^\nu e_{c}{}^\rho \partial_{[\mu} b_{\nu\rho]} \,.
\end{align}
One thus sees that the torsion components $e_{a}{}^\mu e_{b}{}^\nu e_{c}{}^\rho T^{(b)}_{\mu\nu\rho}$ cannot be absorbed in expressions for the spin connections. These can be viewed as constituting another form of intrinsic torsion in SNC geometry and give an extra geometric constraint on the curl of the $b_{\mu \nu}$ field.\footnote{Note that we use the term "torsion" here in the Cartan formulation sense, as the covariantized curl of a frame field. In particular, this type of intrinsic torsion is different from the one considered in section 2, in the sense that it does not correspond to torsion of the affine connection.}

Note that, even after the introduction of the extra torsion equation \eqref{eq:Tisdb}, we have not been able to express all spin connection components in terms of frame fields and torsion. In particular, the following $2(D-2)$ components of $\omega_\mu{}^{A a}$
\begin{align}
  \label{eq:indepcomp}
  \tau_{\{A|}{}^\mu \omega_{\mu|B\}}{}^{a} \equiv \tau_{(A|}{}^\mu \omega_{\mu|B)}{}^{a} - \frac12 \eta_{AB} \tau_C{}^\mu \omega_\mu{}^{C a} \,,
\end{align}
remain independent in our formalism. The full expressions for $\omega_\mu{}^{ab}$ and $\omega_\mu{}^{A a}$ that can be obtained from \eqref{eq:torsionVielbSNC2} and \eqref{eq:Tisdb} are given by
\begin{align}
  \label{eq:fullrotboostomega}
  \omega_\mu{}^{ab} &= -2 e^{[a|\nu|} \partial_{[\mu} e_{\nu]}{}^{b]} + e_{\mu c} e^{a\nu} e^{b\rho} \partial_{[\nu} e_{\rho]}{}^{c} - \frac32 \epsilon_{AB} \tau_\mu{}^A \tau^{B\nu} e^{a\rho} e^{b\sigma} \partial_{[\nu} b_{\rho\sigma]} \nonumber \\ & \qquad + e^{[a|\nu} T_{\mu\nu}{}^{|b]} - \frac12 e_\mu{}^{c} e^{a\nu} e^{b\rho} T_{\nu\rho c} + \frac12 \epsilon_{AB} \tau_\mu{}^A \tau^{B\nu} e^{a\rho} e^{b\sigma} T^{(b)}_{\nu\rho\sigma} \,, \nonumber \\
  \omega_\mu{}^{A a} &= - \tau^{A\nu} \partial_{[\mu} e_{\nu]}{}^{a} + e_{\mu b} \tau^{A\nu} e^{a\rho} \partial_{[\nu} e_{\rho]}{}^{b} + \frac32 \epsilon^{AB} \tau_B{}^\nu e^{a\rho} \partial_{[\mu} b_{\nu\rho]} + \frac12 \tau_\mu{}^B \tau_B{}^\nu \tau^{A\rho} T_{\nu\rho}{}^{a} \nonumber \\ & \qquad - e_{\mu b} \tau^{A\nu} e^{(a|\rho} T_{\nu\rho}{}^{|b)} - \frac14 \tau_\mu{}^A \epsilon_{CD} \tau^{C\nu} \tau^{D\rho} e^{a\sigma} T^{(b)}_{\nu\rho\sigma} - \frac12 e_{\mu b} \epsilon^A{}_B \tau^{B\nu} e^{a\rho} e^{b\sigma} T^{(b)}_{\nu\rho\sigma} \nonumber \\ & \qquad + \tau_\mu{}^B W_B{}^{A a} \,,
\end{align}
where $W_{BA a} = \tau_{\{B|}{}^\mu \omega_{\mu|A\} a}$ corresponds to the independent spin connection components \eqref{eq:indepcomp}.

In order to give the explicit expression for the affine connection, we use a notation to denote certain torsion components:
\begin{align}
  {T_{AB}}^C = {\tau_A}^\mu {\tau_B}^\nu {T_{\mu\nu}}^C \,, \qquad \qquad {T_{Aa}}^B  = {\tau_A}^\mu{e_{a}}^\nu {T_{\mu\nu}}^B \,,\qquad \qquad {T_{ab}}^A = {e_{a}}^\mu{e_{b}}^\nu {T_{\mu\nu}}^A
\end{align}
and similarly for components of $T_{\mu\nu}{}^{a}$ and $T^{(b)}_{\mu\nu\rho}$. This is analogous to what we have used in subsection \ref{ssec:specialTNC}. Plugging the expressions \eqref{eq:exprso11conn} and \eqref{eq:fullrotboostomega} for the spin connections $\omega_\mu$, ${\omega_\mu}^{Aa}$ and ${\omega_\mu}^{ab}$ into eq.~\eqref{eq:GammaVielbpost} leads to the following expression for the affine connection:
\begin{align}
\label{eq:Gammafinalstring}
\Gamma_{\mu\nu}^{\rho}&=
\frac{1}{2}\tau^{\rho\sigma}(\partial_{\mu}\tau_{\nu\sigma}+\partial_{\nu}\tau_{\mu\sigma}-\partial_{\sigma}\tau_{\mu\nu}) +
\frac{1}{2}h^{\rho\sigma}(\partial_{\mu}h_{\nu\sigma}+\partial_{\nu}h_{\mu\sigma}-\partial_{\sigma}h_{\mu\nu})\,
\nonumber\\[.1truecm]
&-\tau_{\mu}{}^{A}e_{\nu}{}^{a}\tau^{\rho B}T_{a(AB)}+\frac12\,e_{\mu}{}^{a}e_{\nu}{}^{b}\tau_C{}^{\rho}T_{ab}{}^\, \nonumber\\[.1truecm]
&+\tau_{\mu}{}^{A}W_{A}{}^{Bc}\tau_{\nu B}e^{\rho}{}_{c} +
\tau_{\nu}{}^{B}\tau^{\rho A}(\tau_{\mu C}T^{C}_{AB}+e_{\mu}{}^{c}T_{c[AB]}) +
\frac{1}{2}e^{\rho}{}_{a}T_{\mu\nu}^{a}-e^{\rho a}e_{b(\mu}T_{\nu)a}^{b}\,\nonumber\\[.1truecm]
&+\frac{1}{4}\tau_{\mu\nu}e^{\rho c}\epsilon^{AB}(3\tau^{\eta}{}_{A}\tau^{\xi}{}_{B}e^{\sigma}{}_{c}\partial_{[\eta}b_{\xi\sigma]}-T^{(b)}_{ABc})\, \nonumber\\[.1truecm]
&-\epsilon^{AB}e^{\rho d}e_{(\mu}{}^{c}\tau_{\nu)A}(
3\tau^{\eta}{}_{B}e^{\xi}{}_{c}e^{\sigma}{}_{d}\partial_{[\eta}b_{\xi\sigma]}
-T^{(b)}_{Bcd})\,,
\end{align}
We do not give an expression in the similar vein as equation \eqref{eq:Gammaexpr2}, as we cannot find a boost covariant quantity that is analogous to $\bar{h}_{\mu\nu}$. Note that the spin connections \eqref{eq:fullrotboostomega} and the affine connection \eqref{eq:Gammafinalstring} are manifestly invariant under the one-form symmetry \eqref{eq:oneformsymm}, if one assumes that $T^{(b)}_{\mu\nu\rho}$ is.

In going from particles to strings, we see that effectively the central charge gauge field $m_\mu$ has been replaced by the 2-form field $b_{\mu\nu}$ which plays a very similar role as $m_\mu$. Both are geometric fields that transform under boost transformations and both are needed to define a dependent spin connection or affine connection that transforms in the correct way. Moreover, $b_{\mu\nu}$ plays a similar physical role as $m_\mu$. Indeed, $b_{\mu\nu}$ acts as a gauge field for the one-form symmetry \eqref{eq:oneformsymm}. One can thus couple it to an anti-symmetric two-tensor current that implements conservation of the string tension via an appropriate Wess-Zumino term \cite{Bidussi:2021ujm}, in analogy to how $m_\mu$ couples to the Noether current corresponding to particle mass conservation.

\subsection{Special cases and examples} \label{ssec:specialTSNC}

\begin{figure}[htp]
\begin{subfigure}[t]{0.35\textwidth}
\centering
\begin{tikzpicture}[scale=0.6,every node/.style={scale=0.6}]
\rectcol{orange} (orange) at (0,2) {\Large $\check{T}^{(b)}_{\mu\nu\rho}$};
\rectcol{blue} (blue) at (-3,-2) {\Large ${T_{\mu\nu}}^{a}$};
\rectcol{purple} (purple) at (3,0) {\Large $T^{(b)}_{abc} $};
\rectcol{green} (green) at (-1,-4) {\Large ${\check{T}_{\mu\nu}}^A$};
\rectcol{yellow} (yellow) at (1,-6) {\Large ${T_{a}}^{(AB)}$};
\rectcol{red} (red) at (3,-8) {\Large ${T_{ab}}^A$};

\draw [-stealth] (blue) to [out= 270,in=180] (yellow);
\draw [-stealth] (blue) to [out= 270,in=180] (red);
\draw [-stealth] (orange) to [out= 0, in=90] (purple);
\draw [-stealth] (orange) to (blue);
\draw [-stealth] (orange) to (green);
\draw [-stealth] (orange) to (yellow);
\draw [-stealth] (purple) to (red);
\draw [-stealth] (blue) to (green);
\draw [-stealth] (green) to (yellow);
\draw [-stealth] (yellow) to (red);


\end{tikzpicture}
\vspace{0.5cm}
\subcaption{This diagram gives a schematic representation of the boost transformations of the different torsion components. Arrows display relations between the different torsion components under boost transformations. If the boost transformation of a torsion component contains another torsion component, an arrow points from the former component towards the latter, e.g. the arrow from the purple to the red box represents the boost transformation $\delta T^{(b)}_{abc} = -3\epsilon_{AB}{\lambda^A}_{[c}{T_{ab]}}^B$.}
\label{fig:stringconsistentboosttransform1a}
\end{subfigure}
\hfill
\begin{subfigure}[t]{0.6\textwidth}
\centering
\begin{tikzpicture}[scale=0.6,every node/.style={scale=0.6}]
\node at (9,0) {$10$};
\rectsix{10}{0}{red}{yellow}{green}{blue}{purple}{orange}

\node at (9,2) {$9$};
\rectsix{10}{2}{white}{yellow}{green}{blue}{purple}{orange}

\node at (5,4) {$7$};
\rectsix{6}{4}{white}{white}{green}{blue}{purple}{orange}

\node at (20,4) {$8$};
\rectsix{14}{4}{white}{yellow}{green}{blue}{white}{orange}

\node at (5,6) {$5$};
\rectsix{6}{6}{white}{white}{white}{blue}{purple}{orange}

\node at (20,6) {$6$};
\rectsix{14}{6}{white}{white}{green}{blue}{white}{orange}

\node at (5,8) {$3$};
\rectsix{6}{8}{white}{white}{white}{white}{purple}{orange}

\node at (20,8) {$4$};
\rectsix{14}{8}{white}{white}{white}{blue}{white}{orange}

\node at (9,10) {$2$};
\rectsix{10}{10}{white}{white}{white}{white}{white}{orange}

\node at (9,12) {$1$};
\rectsix{10}{12}{white}{white}{white}{white}{white}{white}

\draw [-stealth] (12.5,11.5) -- (12.5,10.5);
\draw [-stealth] (12.5,9.5) -- (16.5,8.5);
\draw [-stealth] (12.5,9.5) -- (8.5,8.5);
\draw [-stealth] (16.5,7.5) -- (16.5,6.5);
\draw [-stealth] (16.5,7.5) -- (8.5,6.5);
\draw [-stealth] (8.5,7.5) -- (8.5,6.5);
\draw [-stealth] (16.5,5.5) -- (16.5,4.5);
\draw [-stealth] (16.5,5.5) -- (8.5,4.5);
\draw [-stealth] (8.5,5.5) -- (8.5,4.5);
\draw [-stealth] (16.5,3.5) -- (12.5,2.5);
\draw [-stealth] (8.5,3.5) -- (12.5,2.5);
\draw [-stealth] (12.5,1.5) -- (12.5,0.5);

\end{tikzpicture}
\vspace{0.1cm}
\subcaption{The colours in every rectangle of this diagram represent the parts of the corresponding torsion tensors that are unconstrained. In particular, if a colour is missing, it implies that this part of the torsion tensor is zero. In Figure \ref{fig:stringconsistentboosttransform1a}, we have indicated what colour corresponds to what part of the torsion tensors. Furthermore, arrows point from more special cases towards more general cases.}
\label{fig:stringconsistentboosttransform1b}
\end{subfigure}
\caption{Classification of constraints on the torsion tensors (b) that are consistent with the local structure group transformations (a).}
\label{fig:stringconsistentboosttransform}
\end{figure}
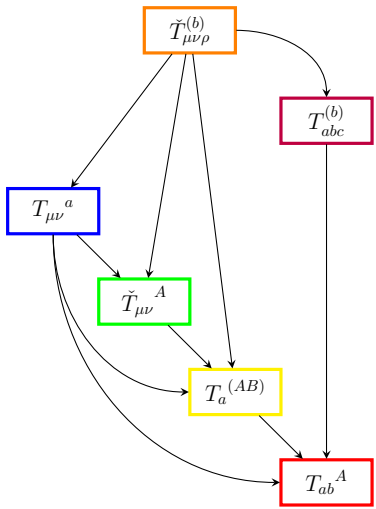
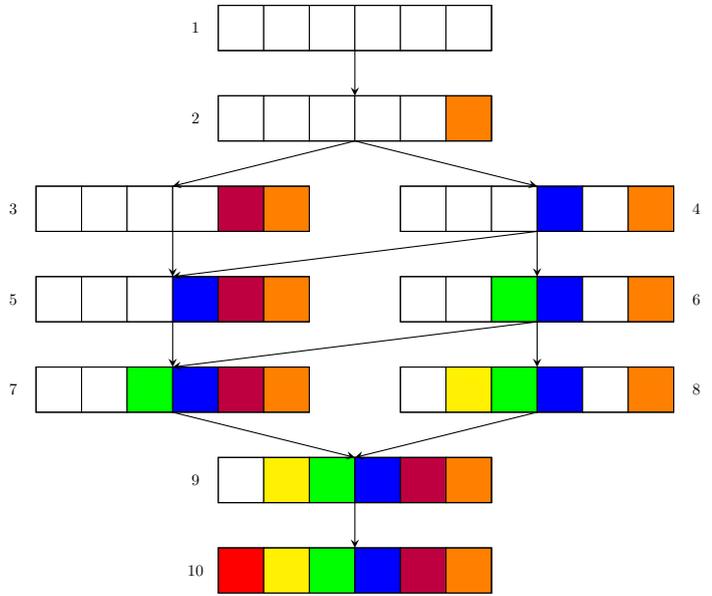

\noindent Similarly as in the particle case in subsection \ref{ssec:specialTNC}, a generic torsionful affine connection that is compatible with SNC geometry includes the torsion tensors ${T_{\mu\nu}}^A$, ${T_{\mu\nu}}^{a}$ and $T^{(b)}_{\mu\nu\rho}$. Those torsion tensors transform under Lorentz transformations, spatial rotations and String Galilean boosts. Some components of the torsion tensors transform to other components, and hence, those torsion tensors cannot be set to zero independently from other torsion components. All possible scenarios in which components of the torsion tensors can be set to zero consistently are displayed in Figure \ref{fig:stringconsistentboosttransform1b}. The structure of this figure is similar to Figure \ref{fig:particleconsistentboosttransform1a}, \ref{fig:particleconsistentboosttransform1b}.

In the following, we find it useful to define the following notation which separates the intrinsic torsion components by defining:
\begin{subequations}
	\begin{align}
		{\check{T}_{\mu\nu}}^A &= {T_{\mu\nu}}^A -  2{e_{[\mu}}^{a}{\tau_{\nu]}^B} \eta_{BC}{T_{a} }^{(AC)} - {e_\mu}^{a} {e_\nu}^{b} {T_{ab}}^A\,,\\
		\check{T}^{(b)}_{\mu\nu\rho} &= T^{(b)}_{\mu\nu\rho} - {e_\mu}^{a}{e_\nu}^{b}{e_\rho}^{c} T^{(b)}_{abc}.
	\end{align}
\end{subequations}
The torsion components ${\check{T}_{\mu\nu}}^A$ and $\check{T}^{(b)}_{\mu\nu\rho}$, are the torsion components ${T_{\mu\nu}}^A$ and $T^{(b)}_{\mu\nu\rho}$ but with the intrinsic torsion projected out. These intrinsic torsion components are given by \eqref{eq:intrinsictorsionSNC} and \eqref{eq:intrinsicdb}.

\begin{figure}[t]
    \centering
    \includegraphics[width=.8\textwidth]{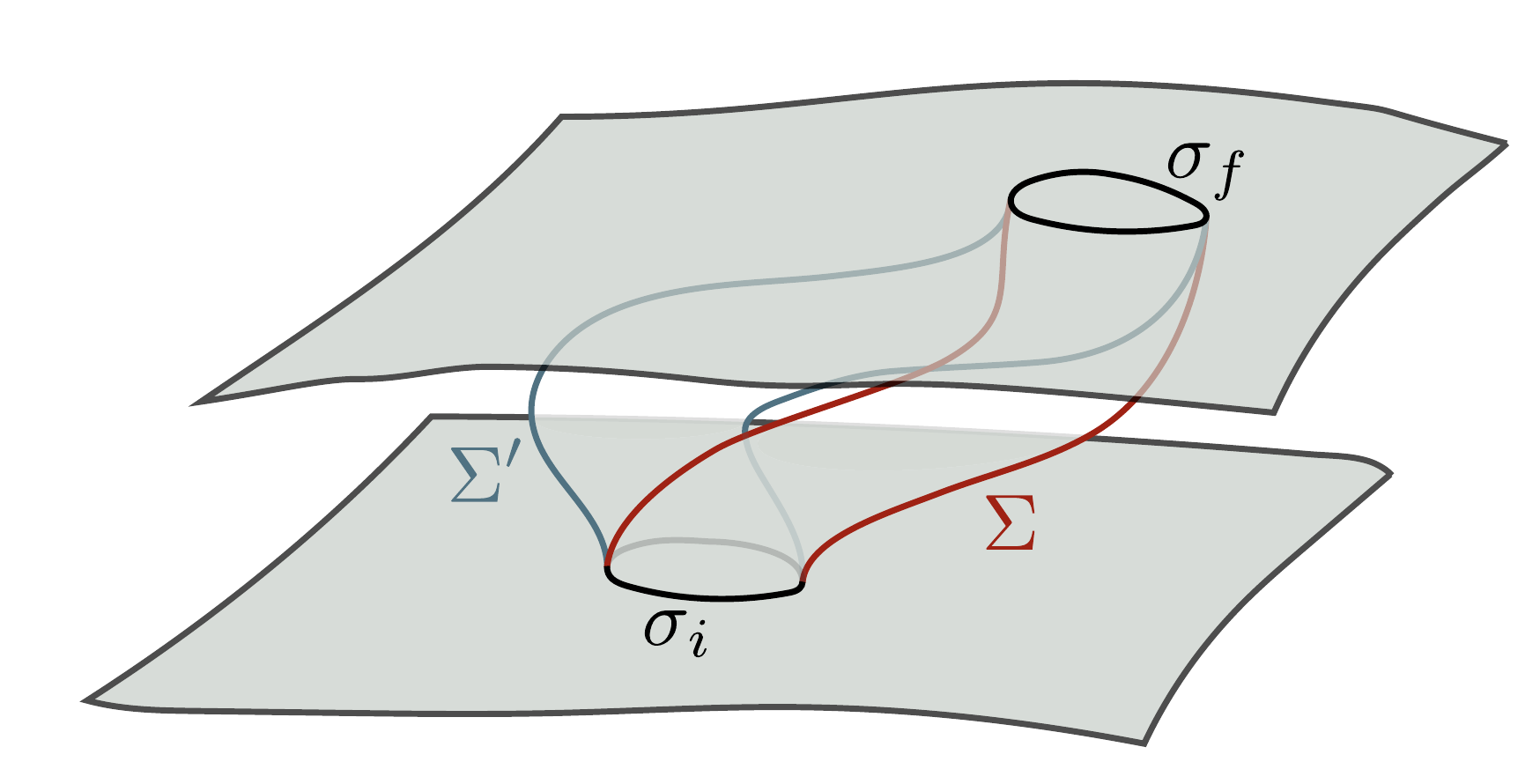}
    \caption{Two worldsheets $\Sigma$ and $\Sigma'$ stretching between two leaves of the manifold with $\partial\Sigma=\partial\Sigma'$. For geometries with zero time-like torsion $T_{\mu\nu}^A=0$, the volume two-form $\ell=1/2\,\ell_{\mu\nu}dx^\mu\wedge dx^\nu$ is closed. We can thus conclude that the proper area \eqref{eq:properarea2} traced out by $\Sigma$ and $\Sigma'$ is the same, i.e. $\int_\Sigma \ell = \int_{\Sigma'} \ell$.}
    \label{fig:AbsoluteArea}
\end{figure}

Analogously to the particle in subsection \ref{ssec:specialTNC}, it is convenient to subdivide those cases. We will do this in the following list.
\begin{itemize}
    \item In cases 1, 2 and 3, we have that the anti-symmetrization of the affine connection is zero. By \eqref{eq:torsiondecompSNC}, this is equivalent to setting ${T_{\mu\nu}}^A=0$ and ${T_{\mu\nu}}^{a}=0$. Those cases are commonly referred to as `torsionless String Newton-Cartan geometry'.
    \item Cases 1, 2, 3, 4 and 5 all have zero longitudinal torsion ${T_{\mu\nu}}^A=0$. In case 4 and 5, we let ${T_{\mu\nu}}^{a}$ unconstrained. In all those cases, the two-form $\ell_{\mu\nu}$ as defined in \eqref{eq:properarea2} is closed, that is $3\,\partial_{[\mu}\ell_{\nu\rho]}=0$. By Stokes' theorem, we obtain that the proper area \eqref{eq:properarea2} is independent of the chosen worldsheet segment $\Sigma_\varphi$ and only depends on the initial and final position of the string, that is, on the curves $\varphi(0,\cdot)=\sigma_i$ and $\varphi(1,\cdot)=\sigma_f$. This implies that the same amount of proper area has been swept out by two strings starting and ending at the same position, irrespective of the worldsheet segment they have traced out throughout space-time. See figure \ref{fig:AbsoluteArea}. This can be rephrased as the statement that SNC manifolds with zero longitudinal torsion admit an absolute area function.
    \item Cases 6 and 7 correspond to setting ${T_{a}}^{(AB)}=0$ and ${T_{ab}}^A=0$ and letting  ${T_{\mu\nu}}^{a}$ unconstrained.
    \item Cases 8 and 9 correspond to setting ${T_{ab}}^A=0$ and letting ${T_{\mu\nu}}^{a}$ and ${T_{a}}^{(AB)}$  unconstrained. In the cases 6, 7, 8 and 9, there does not exist an absolute area function anymore. This means that the area of the worldsheet between two events does not only depend on the initial and final positions of the string, but also on the worldsheet segment a string traces out. As ${T_{ab}}^A=0$, though, there is still a notion of absolute transversal simultaneity. The condition ${T_{ab}}^A=0$ is, by the Frobenius theorem, equivalent to stating that there is a foliation of $(D-2)$ transversal submanifolds, i.e. submanifolds of the space-time manifold $\mathcal{M}$ such that the tangent vectors to all curves on those submanifolds are transversal, as defined in subsection \ref{ssec:framemetricTSNC}. A notion of string causality that distinguishes between past and future can be defined as the following statement: a string defined by the embedding $\sigma_f: [0,1] \to \mathcal{M}$ is in the future with respect to a string defined by an embedding $\sigma_i: [0,1] \to \mathcal{M}$ if there exists a worldsheet segment $\Sigma_\varphi$ with $\varphi(0,\cdot)=\sigma_i$ and $\varphi(1,\cdot)=\sigma_f$ such that the integral in \eqref{eq:properarea2} is positive.
    %
    \item Case 10 is generic torsion. As we let ${T_{ab}}^A$ unconstrained, consistency with boost transformations requires that all other torsion components are also unconstrained. There is no notion of absolute area or absolute transversal simultaneity anymore.
\end{itemize}

The above classification needs a further refinement if we include the intrinsic torsion constraints that describe the DSNC${}^-$ geometry underlying non-relativistic string theory with $N=1$ supersymmetry \cite{Bergshoeff:2021tfn}, since these constraints set part of the torsion tensors
\begin{equation}
T_{ab}{}^A\hskip 1truecm \textrm{and} \hskip 1truecm T_{aA}{}^B
\end{equation}
equal to zero without changing the basic structure of the classification. This proceeds in two steps. First one picks out those intrinsic torsion components that are invariant under local (anisotropic) dilatations  $\delta \tau_\mu{}^A = \lambda_D\tau_\mu{}^A$, since they are an emerging symmetry in non-relativistic string theory. One thus ends up with the components
\begin{equation}\label{d}
T_{ab}{}^A\hskip 1truecm \textrm{and} \hskip 1truecm T_{a\{AB\}}\,,
\end{equation}
where $\{AB\}$ stands for the symmetric traceless part of $A$ and $B$. Formally, these tensors can be obtained by discarding $T_{aA}{}^A$. Since it transforms as a (dependent) dilatation gauge field, it should not be seen as part of the intrinsic torsion of the geometry. In a second step, using light-cone notation $A=(+,-)$, we set half of the intrinsic torsion components given in \eqref{d} equal to zero:
\begin{equation}\label{d2}
T_{ab}{}^- = T_{a+}{}^- = 0\,.
\end{equation}
One then obtains the DSNC${}^-$ case by setting these torsion components together with $T_{aA}{}^A$ to zero in the above classification.

Let us now give an example of dependent torsion tensors, similar to what we considered at the end of section \ref{ssec:specialTNC}. We assume that we are in a dimension where there exists a vector-spinor $\psi_\mu$ that satisfies the Majorana condition. A concrete example in ten dimensions has been worked out in detail in \cite{Bergshoeff:2021tfn}. The vector-spinor forms a representation of the local $\mathrm{SO}(1,1)\times\mathrm{SO}(D-2)$ transformations
\begin{align}\label{eq:stringspinortrans}
  \delta\psi_\mu = -\frac12\,\lambda_M\,\gamma_{01}\psi_\mu + \frac14\,\lambda^{ab}\gamma_{ab}\psi_\mu\,,
\end{align}
where the gamma matrices $(\gamma_A,\gamma_{a})$ form a Clifford algebra with signature $(-++\cdots+)$. In order to specify the boost representation, it is useful to split the spinor as $\psi_\mu = \psi_{\mu+} + \psi_{\mu-}$, where the components are eigenspinors under $\gamma_{01}=\gamma_0\gamma_1$ as follows: $\gamma_{01}\psi_{\mu\pm}=\pm\psi_{\mu\pm}$. Equivalently, one can define $\psi_{\mu\pm}=1/2(\mathds{1}\pm\gamma_{01})\psi_\mu$. The transformation under String Galilean boosts with parameters $\lambda^{Aa}$ is then given as
\begin{align}\label{eq:stringspinorboost}
    &\delta\psi_{\mu+}=0\,,&& \delta\psi_{\mu-} = \frac12\,\lambda^{Aa}\gamma_{A}\gamma_{a}\psi_{\mu+}\,.
\end{align}
The projected Majorana spinors are also the natural building blocks for constructing independent spinor bilinears as follows:
\begin{align}\label{eq:fermbiltor1}
  &T_{\mu\nu}{}^A = \frac12\,\bar\psi_{[\mu +}\gamma^A\psi_{\nu]+}\,, && T_{\mu\nu}{}^{a} = \bar\psi_{[\mu+}\gamma^{a}\psi_{\nu]-}\,.
\end{align}
Using the transformation rules \eqref{eq:stringspinortrans} and \eqref{eq:stringspinorboost}, one can then show that the two-forms $T_{\mu\nu}{}^A$ and $T_{\mu\nu}{}^{a}$ transform as given in equation \eqref{eq:trafoTATAp}. Due to the identity $\epsilon^{AB}\gamma_B = -\gamma^A\gamma_{01}$ and the properties of the projected spinors, we find that $\epsilon^A{}_{B}T_{\mu\nu}{}^B = -T_{\mu\nu}{}^A$. This is equivalent to the statement that $T_{\mu\nu}{}^- = 2^{-1/2}(T_{\mu\nu}{}^0-T_{\mu\nu}{}^1)=0$ identically. The three-form torsion can analogously be defined as
\begin{align}\label{eq:fermbiltor2}
  &T^{(b)}_{\mu\nu\rho} = 3\,\bar\psi_{[\mu-}\gamma^A\psi_{\nu-}\tau_{\rho]A} -3\,T_{[\mu\nu}{}^{a}e_{\rho]a}
\end{align}
It is straightforward to check that this three-form transforms under the local structure group as in \eqref{eq:boostT}. The tensors given in equations \eqref{eq:fermbiltor1} and \eqref{eq:fermbiltor2} provide explicit examples of the dependent torsion tensors. Consequently, they gives rise to an affine connection that is invariant under String Galilean boosts.

\section{Conclusions and Outlook} \label{sec:conclusions}

\noindent In this work we gave an in-depth description of generalized NC geometries for particles and strings using a frame formulation. In the case of particles, such a frame formulation stresses the relation with the underlying structure group and makes it possible to derive several results in an elegant way.  An important feature of our discussion was the introduction of independent torsion tensors which makes it possible to define spin connections and affine connections that transform in the right way under the symmetries of the structure group. We gave a rather extensive set of solutions of different constraints that one can impose on the intrinsic torsion tensors leading to different constrained geometries. Furthermore, we gave a physical interpretation of the geometric fields at several places, thereby extending the notion of absolute time to the string case.

One might wonder whether there is a natural interpretation of the 2-form field $b_{\mu\nu}$ similar to the interpretation of the gauge field $m_\mu$ as the one associated with the central extension of the Galilei algebra. One interesting proposal, inspired by earlier work in supergravity, was recently given in \cite{Bidussi:2021ujm}  where the 2-form field $b_{\mu\nu}$ was represented as a dependent expression in terms of two 1-form gauge fields. A drawback of this description is that the reducible gauge symmetry of $b_{\mu\nu}$ cannot be mimicked by the irreducible gauge symmetries of the 1-form gauge fields. Furthermore, representing $b_{\mu\nu}$ as the product of two gauge fields implies the existence of additional Stueckelberg symmetries whose algebraic origins are not clear. An alternative option is to go to loop space geometry, thereby replacing coordinates $x^\mu$ by $x^\mu(\sigma)$, where the coordinate $\sigma$ parametrizes a circle,  and replacing fields $\phi(x)$ defined over ordinary geometry  by fields $\phi\big(x(\sigma)\big)$  defined over loop space geometry.
Within such a geometry, it  is natural to define a loop space covariant derivative involving the 2-form  $b_{\mu\nu}$ as follows \cite{Bergshoeff:2001kp}:
\begin{equation}
\mathcal{D}_\mu(\sigma) = {\delta\over \delta x^\mu(\sigma)} - b_{\mu\nu}x^{\prime\nu}\,.
\end{equation}
This naturally corresponds to a loop algebra with generators $T(\sigma)$. Although promising, it is not yet clear how useful this approach is. At the moment, perhaps a more practical approach is to  work immediately in terms of fields and ignore a possible relation with an underlying algebra which is not needed at least for the purpose of this work.

In \cite{Figueroa-OFarrill:2020gpr}, the intrinsic torsion of non-Lorentzian geometric structures was systematically studied and classified using cohomological techniques. The classification derived there agrees with the one given in section \ref{sec:TNC}. It would be interesting to see whether the analysis based on Spencer cohomology can be extended to the study of SNC-type geometries as presented in section \ref{sec:TNC2}. Furthermore, it would be natural to generalize that to G-structures with G=$(\mathrm{SO}(1,p)\times\mathrm{SO}(D-p-1))\rtimes\mathds{R}^{(p+1)(D-p-1)}$---that is, so-called $p$-brane geometries \cite{Pereniguez:2019eoq}.

It is natural to consider the extension of our work to non-relativistic string theory with $N=2$ supersymmetry and to M-theory or membranes. In the case of $N=2$ string theory, one expects more constraints than the ones characterizing the DSNC${}^-$ geometry given in equation \eqref{d2}. These will also include fermionic intrinsic torsion tensors. We expect the same to happen for M-theory with the understanding that in that case one uses a membrane foliation \cite{Blair:2021waq,Ebert:2021mfu} with $A=0,1,2$ and $a = 3, \cdots, 10$. This suggests the existence of a degenerated supergeometry whose proper formulation might require the use of superfields and superspace. The non-relativistic torsion constraints we find are reminiscent of the superspace torsion constraints that one imposes in the relativistic case to define a relativistic supergravity theory. Once constructed, by consistency the non-relativistic M-theory geometry one finds should reduce to the DSNC${}^-$ geometry considered in this work by performing a double dimensional reduction over a spatial membrane direction followed by a truncation. We hope to come back to these issues in a forthcoming work.

\section*{Acknowledgements}

\noindent We would like to thank José Figueroa-O'Farrill, Quim Gomis, Julian Kupka, and Roland van der Veen for valuable discussions. KVH is funded by the Fundamentals of the Universe program at the University of Groningen. The work of LR has been initially supported by the FOM/NWO free program Scanning New Horizons and successively supported by Next Generation EU through the Maria Zambrano grant from the Spanish Ministry of Universities under the Plan de Recuperacion, Transformacion y Resiliencia.

%

\providecommand{\href}[2]{#2}\begingroup\raggedright\endgroup

\end{document}